\documentclass{emulateapj}
\usepackage{graphicx}
\usepackage{natbib}
\usepackage{multirow}
\usepackage{afterpage}
\def\Lsol {$\hbox{L}_\odot$}

\def\Vlsr {V$_{\rm LSR}$}
\def\Msol {$\hbox{M}_\odot$}

\def\kms {$\rm km~s^{-1}$}
\def\klambda {k$\lambda$}

\slugcomment{Accepted for publication in ApJ}

\shorttitle{Orion Source I}
\shortauthors{Plambeck \& Wright}

\begin{document}

\title{ALMA Observations of Orion Source I at 350 and 660 GHz}

\author{R.~L.~Plambeck, M.~C.~H.~Wright}
\affil{Radio Astronomy Laboratory, University of California, Berkeley, CA 94720, USA}
\email{plambeck@berkeley.edu}

\begin{abstract}

Orion Source I (``SrcI'') is the protostar at the center of the Kleinmann-Low
Nebula.  ALMA observations of SrcI with 0.2\arcsec\ angular resolution were
made at 350 and 660~GHz to search for the H26$\alpha$ and H21$\alpha$ hydrogen
recombination lines and to measure the continuum flux densities.  The
recombination lines were not detected, ruling out the possibility that SrcI is
a hypercompact HII region.  The deconvolved size of the continuum source is
approximately 0\farcs23 $\times$ 0\farcs07 ($\sim 100\times30$~AU);
it is interpreted as a disk viewed almost edge-on.  Optically thick thermal
emission from $\sim$500~K dust is the most plausible source of the continuum,
even at frequencies as low as 43~GHz; the disk mass is most likely in the range
0.02-0.2~\Msol.  A rich spectrum of molecular lines is detected, mostly from
\mbox{sulfur-} and silicon-rich molecules like SO, SO$_2$, and SiS, but also
including vibrationally excited CO and several unidentified transitions.  Lines
with upper energy levels E$_{\rm U} > 500$~K appear in emission and are
symmetric about the source's LSR velocity of 5~\kms, while lines with E$_{\rm
U} < 500$~K appear as blueshifted absorption features against the continuum,
indicating that they originate in outflowing gas.  The emission lines exhibit a
velocity gradient along the major axis of the disk that is consistent with
rotation around a 5-7~\Msol\ central object.   The relatively low mass of SrcI
and the existence of a 100~AU disk around it are difficult to reconcile
with the model in which SrcI and the nearby Becklin-Neugebauer Object were
ejected from a multiple system 500 years ago.

\end{abstract}

\vspace{6pt} 


\section{INTRODUCTION}
\label{sec:intro} 

The Kleinmann-Low Nebula in Orion, at a distance of 415~pc
\citep{Menten2007,Kim2008}, is the nearest region in which high mass stars (M
$>8$~\Msol) are forming. Much of the $\sim 10^5$~\Lsol\ luminosity
\citep{Werner1976} of the region originates from a deeply embedded object,
radio Source~I (hereafter, ``SrcI'').  Although SrcI is so heavily obscured by
foreground dust that it cannot be seen directly at infrared wavelengths, light
reflected off the surrounding nebulosity provides a glimpse of its near
infrared spectrum \citep{Morino1998}.  \citet{Testi2010} infer that the NIR
emission is produced in a disk around a $\sim$10~\Msol\ star, accreting at a
very high rate of a few~$\times$~10$^{-3}$~\Msol~yr$^{-1}$.  

The radio continuum from SrcI has been imaged with the VLA at 43~GHz
\citep{Reid2007,Goddi2011} and with CARMA at 229~GHz \citep{Plambeck2013}.  The
continuum source is elongated, consistent with a 100~AU diameter disk viewed
nearly edge-on.  The nature of the radio emission is uncertain.  SrcI's flux
density is proportional to $\nu^2$ from 43~GHz to 229~GHz, suggesting that the
emission is optically thick over this frequency range, but the brightness
temperature calculated from the high resolution images is $\lesssim$1500~K,
much lower than expected for an HII region.  Although these results do not rule
out the possibility that SrcI is an exceptionally dense HII region that is not
fully resolved by the observations, \citet{Plambeck2013} concluded that the
continuum most likely originated from electron-{\it neutral} free-free emission
(the H$^-$ opacity), as proposed originally by \citet{Reid2007}.  The H$^-$
opacity is important at temperatures below 4500~K where electrons, produced by
ionization of Na, K, and other metals, scatter off neutral hydrogen atoms or
molecules.  If the temperature is as low as 1500~K, a massive ($>$1~\Msol) disk
is required in order for the H$^-$ emission to be optically thick at 229~GHz
\citep{Hirota2015}.

SrcI is one of the few young stars associated with SiO masers.  The maser
spots, mapped in exquisite detail with VLBI observations
\citep{Kim2008,Matthews2010}, are clustered along the top and bottom surfaces
of the continuum disk.  A bipolar outflow also is launched from the disk and
propagates thousands of AU into the surrounding medium \citep{Plambeck2009}.

Although the results described so far are consistent with the formation of SrcI
through disk accretion, as in the models of \citet{McKee2003}, other
observations suggest that stellar interactions played a key role in its
evolution.  In particular, proper motion measurements suggest that SrcI and the
other massive star in this region, the Becklin-Neugbebauer Object (BN), are
recoiling from one another at 35--40~\kms\
\citep{Rodriguez2005,Gomez2008,Goddi2011}.  Tracing the proper motions
backward, \citet{Goddi2011} find that 560 years ago the projected separation of
the two stars was just $50 \pm 100$~AU.  This discovery motivated the currently
popular model in which SrcI and BN were ejected from a multiple system
following a close dynamical interaction \citep{Gomez2008,Bally2011,Goddi2011}.
SrcI is moving at about half the speed of BN, which is known to be a
$\sim$10~\Msol\ B star.  Conservation of momentum therefore suggests that SrcI
has a mass of order 20~\Msol.  This, however, conflicts with the mass of
$\sim$8~\Msol\ derived from SiO maser rotation curves
\citep{Kim2008,Matthews2010}.  It also is surprising that a disk around SrcI
could have survived the dynamical encounter or could have accumulated via
Bondi-Hoyle accretion in just 500 years \citep{Bally2011}.

In this paper we present the results of 0\farcs2 resolution ALMA observations
of SrcI that were designed to determine the nature of its continuum emission
and further constrain its mass.  These observations searched for the
H26$\alpha$ and H21$\alpha$ hydrogen recombination lines, at 353 and 662~GHz,
frequencies high enough that any free-free continuum must be optically thin.
Neither recombination line was detected, ruling out the possibility that the
source is a hypercompact HII region.  The data provide the most reliable
measurements to date of SrcI's flux densities at submm wavelengths, and reveal
a rich spectrum of high excitation (500-3500~K) emission lines from it.
Rotation curves derived from the emission lines are consistent with a central
mass of 5-7~\Msol, lower than predicted by the dynamical ejection model.  

\section{OBSERVATIONS AND DATA REDUCTION}
\label{sec:obs}

Details of the ALMA observations are given in Table~1.  The correlator was set
up in the Frequency Division Mode, with $3840 \times 0.488$~MHz channels
$\times$~2~polarizations in each of 4 spectral windows.  For Band~7, spectral
window 0 was centered at 353.623~GHz, the frequency of the H26$\alpha$
recombination line; for Band~9, spectral window 0 was centered close to
662.404~GHz, the H21$\alpha$ frequency.  The remaining spectral windows were
configured to observe frequency ranges that were relatively free of strong
spectral lines in the surveys of \citet{Schilke1997} and \citet{Schilke2001}.
An angular resolution of better than 0\farcs25 was requested in order to
cleanly resolve SrcI from the adjacent Orion hot core.  The pointing center
was $05^{\rm h}35^{\rm m}14\fs514$, $-5\degr22\arcmin30\farcs56$.

% ============================= Table 1 ==========================
\begin{deluxetable}{lll} 
\tablecaption{ALMA Observations of Orion-KL}
\tablehead{  & \colhead{Band 7} & \colhead{Band 9}  }
\startdata 
   date 					& 2014-07-26 			& 2014-07-06 \\
   no. of antennas 			& 31 					& 28 \\
   precip water 			& 0.32~mm 				& 0.14~mm \\
   on-source time   		& 24~min 				& 15.6 min \\
   uvrange 					& 31--733 m  			& 19--630 m \\
     						& (35--860 k$\lambda$)	& (40--1390 k$\lambda$) \\
   spw0 					& 352.7--354.6~GHz 		& 661.3--663.2 GHz \\
   spw1 					& 354.5--356.4~GHz 		& 663.2--665.1 GHz \\
   spw2 					& 340.6--342.4~GHz 		& 665.9--667.8 GHz \\
   spw3 					& 342.4--344.3~GHz 		& 649.3--651.2 GHz \\
   spectral resolution 		& 0.84~\kms 			& 0.44~\kms \\
   primary beamwidth    	& 16\farcs5 FWHM 		& 9\farcs1 FWHM 
\enddata 
\end{deluxetable} 
% ============================= Table 1 ==========================

The raw visibility data were calibrated using ALMA-supplied scripts and the
CASA software package.  The calibrated CASA measurement sets then were written
out in FITS format and imported into {\tt Miriad}\footnote{{\tt Miriad}
variable {\tt veldop}, the observatory velocity relative to the LSR reference
frame, is not included in the FITS file; it is filled in by specifying keyword
{\tt velocity=lsr} when importing the data with {\tt Miriad} task {\tt fits}.}.
All further processing was done with {\tt Miriad}.

% ============================= Figure spw2  ==========================
\begin{figure} \centering
% trim left bottom right top
\includegraphics[width=1.0\columnwidth, clip, trim=0.5cm 0cm 1cm 1cm] {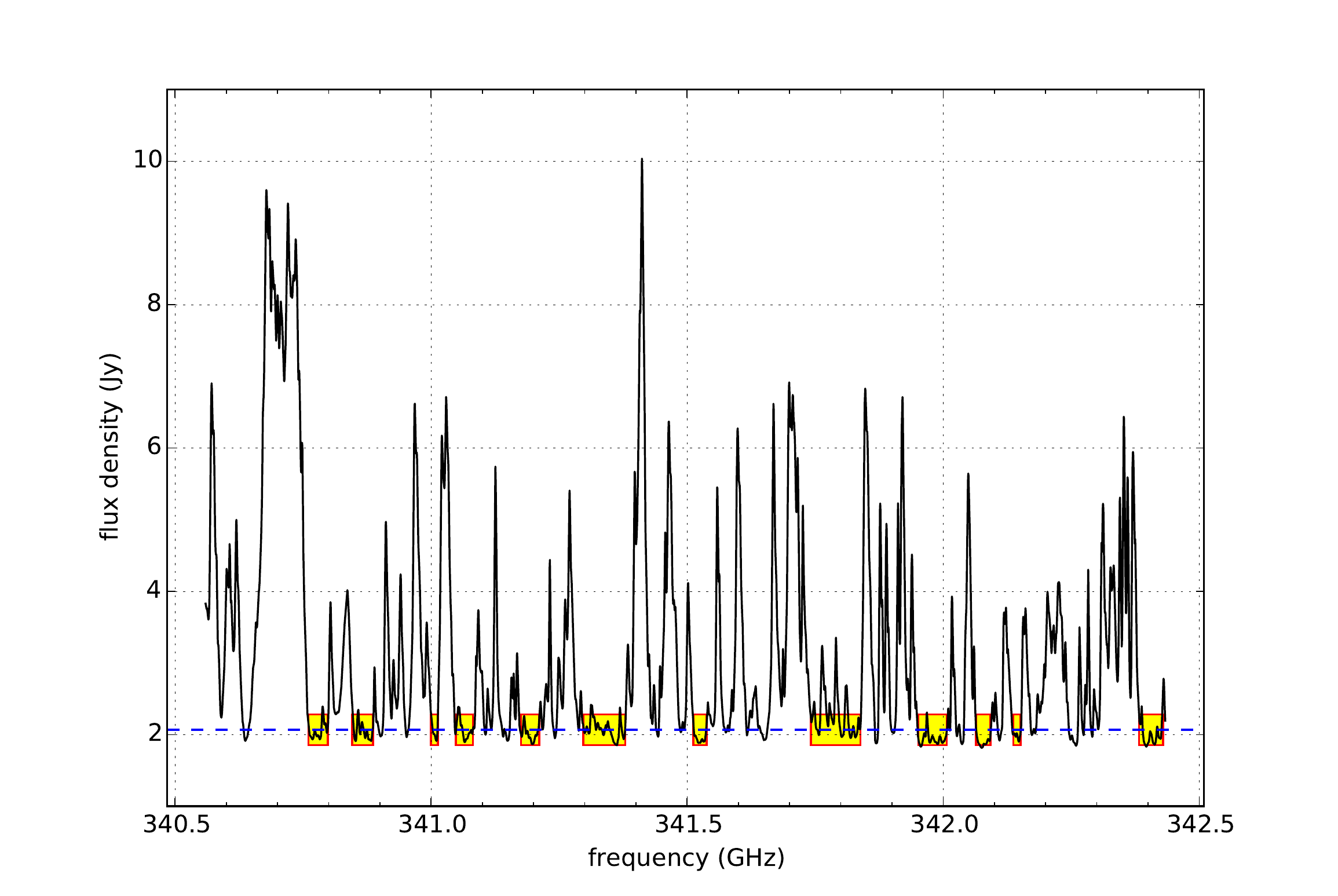} 

\caption{Spectrum of Orion-KL in spectral window 2 of Band~7, illustrating the
difficulty of finding ``line free'' spectral channels.  Shaded boxes indicate
the spectral ranges that were included in the continuum map.  This spectrum is
the {\it scalar} average of the interferometric visibility amplitudes obtained
on projected baselines shorter than 200~\klambda; it differs from a single dish
total power spectrum because it is insensitive to emission that is uniform on
spatial scales $>2$~\arcsec.}

\label{fig:spw2} 
\vspace{6pt} 
\end{figure}
% ============================= Figure spw2  ==========================

% ============================= Figure uvcov  ==========================
\begin{figure*} \centering
% trim left bottom right top
\includegraphics[width=0.8\textwidth, clip, trim=1cm 6cm 2cm 7cm] {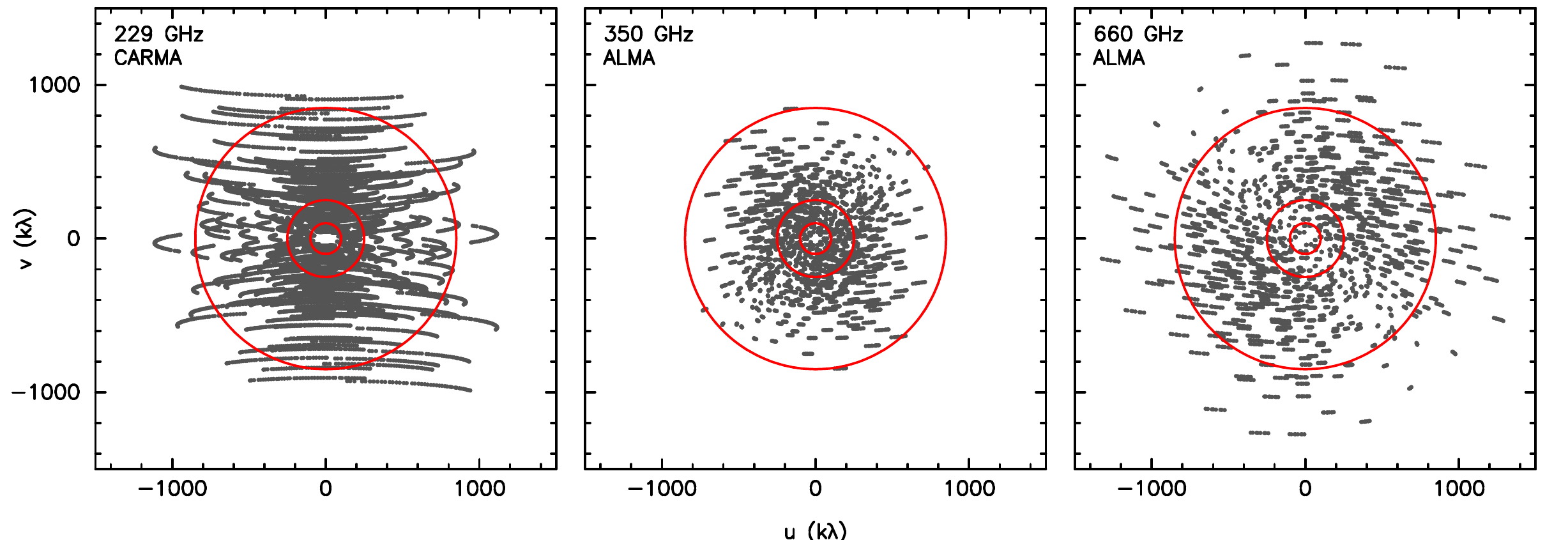} 

\caption{$(u,v)$ coverage of the data.  The 229 GHz data are from the CARMA A and B
arrays \citep{Plambeck2013}.  Red circles indicate radii of 100, 250, and
850~k$\lambda$.}

\label{fig:uvcov} 
\vspace{12pt} 
\end{figure*}
% ============================= Figure uvcov  ==========================

\subsection{Identifying Line-Free Channels}
\label{sec:line-free}

To produce continuum images it is necessary to omit spectral channels that are
contaminated by molecular line emission, a challenging task for a chemically
rich region such as Orion.  We identified such channels by plotting the {\it
scalar} average of the interferometric visibility data.  An example is shown in
Figure~\ref{fig:spw2}.  High amplitudes in this spectrum indicate that the
source has brightness variations on scales of 0\farcs5--2\arcsec\ due to the
presence of molecular lines.  We note that this average does not contain total
power data, so it is not the same as the spectrum that one would observe with a
single dish; for example, an optically thick line with relatively uniform
brightness across the primary beam would not be prominent in the scalar
visibility average.

It was more difficult to identify line-contaminated channels in Band~9 using
this technique because ozone lines in the earth's atmosphere increase the
system noise temperature, producing broad, cuspy peaks in the spectral noise
floor.  We note that Band~9 also uses double sideband receivers, so that ozone
lines in either sideband generate these noise peaks, although astronomical
signals from only one sideband are present.

We adjusted the list of line-contaminated channels after generating a
preliminary set of channel images and inspecting the spectrum of SrcI.

\subsection{Self-Calibration}
\label{sec:selfcal}

After flagging the line-contaminated channels, we averaged the remaining
channels in each spectral window into a single channel for continuum imaging.
A separate image was generated for each spectral window, omitting visibility
data from projected baselines shorter than 250~k$\lambda$ (corresponding to a
fringe spacing of 0.82\arcsec) in order to filter out extended
emission\footnote{Visibilities measured on short spacings represent low spatial
frequency Fourier components of the source brightness distribution, primarily
thermal dust emission from large scale structures like the Orion Hot Core.  Our
data do not fully sample these short spacings, and the missing Fourier
components introduce ripples into the images.  Omitting all the short spacing
data greatly reduces the magnitudes of these ripples.}.  Data were weighted
using Briggs {\tt robust=0}. The images were dominated by emission from SrcI at
the center.

Two iterations of phase-only self-calibration were then applied to the data.
The images were deconvolved with the {\tt CLEAN} algorithm, then
self-calibrated using {\tt CLEAN} components stronger than 3$\sigma$ as a
source model.  The self-calibration interval was chosen to be 6.05~secs, equal
to the integration time.  Self-calibration corrects the data for atmospheric
phase fluctuations that occur on time scales longer than this.  We judged the
quality of the self-calibration by comparing, for each 6.05-sec integration,
the antenna-based phase corrections derived independently for the 4 spectral
windows.  The {\it mean} phase difference between pairs of bands ranged from 0
to 4\degr; perfect agreement is not expected since atmospheric delays produce
chromatic phase differences.  The rms scatter about the mean ranged from
3\degr\ to 8\degr, which indicates that the signal to noise of the calibration
procedure is high.  Self-calibration increased the flux density of SrcI by
approximately 25\% at 350 GHz and 50\% at 660 GHz.  For both Band 7 and Band 9,
the flux density in each of the 4 spectral windows was within 3\% of the mean,
suggesting that no window was seriously contaminated by molecular line
emission.

We did not use amplitude self-calibration because for some antennas this
yielded unreasonably large (up to 50\%) gain drifts on time scales of 20
minutes, probably because of contributions from poorly sampled large
scale structures to the visibility amplitudes.

\subsection{229 GHz CARMA Data}

To compare continuum flux densities and source sizes over a wide frequency
range, we also make use of 229~GHz Orion observations obtained with the
Combined Array for Research in Millimeter Astronomy (CARMA) A and B arrays in
2009 and 2011.  Maps based on these data were previously published by
\citet{Plambeck2013}, but here we combine the A and B array data to match the
spatial frequency range of the ALMA data as closely as possible.  Details of
the CARMA calibrations may be found in \citet{Plambeck2013}; a 2 minute time
interval was used for the final phase-only self-calibration on Orion.

Figure~\ref{fig:uvcov} compares the {\it (u,v)} coverage (sampling in the spatial
frequency domain) for the 3 datasets.  The CARMA data are observations over
full $(u,v)$ tracks, while the ALMA data are snapshot observations.  The
660~GHz data provide the highest angular resolution; 350~GHz, the lowest.

% ================== Table gfits (fullwidth) =====================
\begin{deluxetable*}{cccccccc}
\tablecaption{Gaussian Fits to SrcI}
\tablehead{\colhead{freq} & \colhead{synth beam} & \colhead{$\alpha$(J2000)} & \colhead{$\delta$(J2000)} & epoch & 
           \colhead{integrated flux} & \colhead{deconvolved size\tablenotemark{a}} & \colhead{P.A.} \\ 
           \colhead{(GHz)} & \colhead{(arcsec, P.A.)} & \colhead{(h m s)} & \colhead{(\degr~\arcmin~\arcsec)} & &
           \colhead{(mJy)} & \colhead{(arcsec)}  & \colhead{(\degr)} }   
\startdata 
\phantom{2}43\tablenotemark{b} 
     & $0.06 \times 0.04$, $3\degr$   & 05 35 14.5141 & -05 22 30.575 & 2009.1 &   $11 \pm 2$   & $0.23 \pm 0.01 \times 0.12 \pm 0.01$ & $142 \pm 2$ \\
229  & $0.18 \times 0.15$, $72\degr$  & 05 35 14.5136 & -05 22 30.584 & 2009.1 &  $330 \pm 50$  & $0.21 \pm 0.02 \times 0.07 \pm 0.03$ & $140 \pm 2$ \\
349  & $0.25 \times 0.18$, $66\degr$  & 05 35 14.5152 & -05 22 30.564 & 2014.6 &  $790 \pm 80$  & $0.24 \pm 0.03 \times 0.07 \pm 0.06$ & $140 \pm 1$ \\
661  & $0.16 \times 0.12$, $-50\degr$ & 05 35 14.5129 & -05 22 30.576 & 2014.5 & $3455 \pm 520$ & $0.25 \pm 0.02 \times 0.08 \pm 0.02$ & $142 \pm 1$ 
\enddata 
\tablenotetext{a}{Mean of sizes fit to the {\tt CLEAN} image, the {\tt MAXEN} image, and directly to the visibility data.}
\tablenotetext{b}{\citet{Goddi2011}.}
\label{tab:gfits}
\end{deluxetable*} 
% ================== Table gfits (fullwidth) =====================

\section{CONTINUUM IMAGES}
\label{sec:continuum}

% ============================= Fig maps3 ==========================
\begin{figure} \centering
% trim left bottom right top
\includegraphics[width=1.0\columnwidth, clip, trim=8cm 2.5cm 7cm 1.1cm] {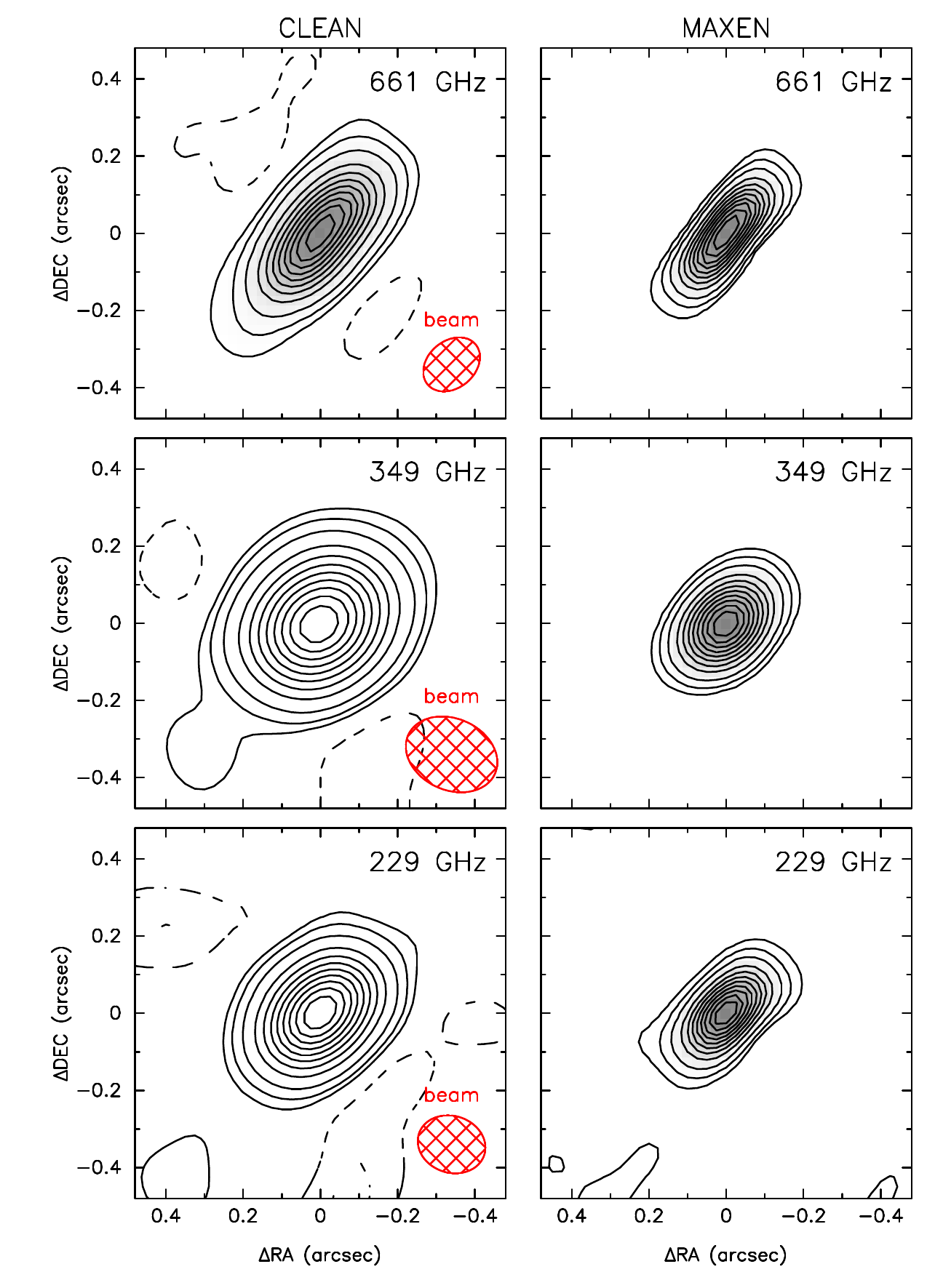} 

\caption{Comparison of SrcI continuum images at 229, 349, and 661~GHz.  The
229~GHz data are from CARMA.  Left-hand panels show images deconvolved with
{\tt CLEAN}; the synthesized beam sizes are given in Table~\ref{tab:gfits}.
Right-hand panels show images that were deconvolved with a maximum entropy
algorithm; these have not been convolved with a restoring beam.  The contour
levels are ($\pm 2.5$, 5, 10, 20, $\dots$ 90\%) of the peak flux densities,
which are 0.19, 0.52, 1.68~Jy/beam (at 229, 349, 661~GHz) for the {\tt CLEAN}
images, and 7.8, 12, and 76~mJy/pixel (0\farcs25 $\times$ 0\farcs25 pixels) for
the {\tt MAXEN} images.}

\label{fig:maps3} 
\vspace{12pt} 
\end{figure}
% ============================= Fig maps3 ==========================

Figure~\ref{fig:maps3} displays continuum images of SrcI centered at
frequencies of 229, 349, and 661~GHz that were generated from the
self-calibrated data on baselines $>$250~\klambda, using {\tt robust=0}
weighting.  The channel-averaged visibilities for the 4 separate spectral
windows were combined using multifrequency synthesis to avoid bandwidth
smearing at the edge of the images.  Each image covered a field of view greater
than the FWHM of the telescope primary beamwidth, but Figure~\ref{fig:maps3}
shows only a 1\arcsec\ box centered on SrcI.  

The images on the left were deconvolved with {\tt CLEAN}, those on the right
with a maximum entropy algorithm ({\tt Miriad} task {\tt MAXEN}).  The {\tt
MAXEN} images are super-resolved in the sense that they have not been convolved
with a restoring beam.  Since {\tt MAXEN} attempts to produce the flattest map
that is consistent with the data, the results are sensitive to the noise level
that is estimated for the data---we used the rms noise measured from the {\tt
CLEAN} residual map.  The flux densities were left unconstrained in the {\tt
MAXEN} images.

To measure SrcI's size and integrated flux density, we fit both Gaussian and
disk models to the emission in the central 0\farcs8 box of the %{\tt CLEAN}
images using {\tt Miriad} program {\tt imfit}.  We judged the quality of each
fit from the rms noise in this box after subtracting the model image. There was
little difference in the residual noise between Gaussian and disk models, so we
discuss only the results of the Gaussian fits here.  Table~\ref{tab:gfits}
summarizes these results.  For comparison, the table also includes the
43~GHz source parameters derived by \citet{Goddi2011} from VLA images.  

SrcI is considerably narrower than the synthesized beams, so our
measurements of the minor axis are particularly uncertain.   For all 3
frequencies we found that deconvolved widths fitted to the {\tt CLEAN} images
were smaller than those measured from {\tt MAXEN} images, or to widths fitted
directly to the $>$250~\klambda\ visibility data with program {\tt
uvfit}\footnote{A disadvantage of fitting a source model directly to the
visibilities is that {\it all} sources within the telescope primary beam
contribute to these visibilities; however, if the data are well-sampled in the
{\it (u,v)} plane, the contributions from sources away from the center of the
beam should cancel when the data are vector-averaged.}. For example, at 229~GHz
the {\tt CLEAN}, {\tt uvfit}, and {\tt MAXEN} minor axis widths were
0\farcs047, 0\farcs072, and 0\farcs105, respectively.  Therefore, in
Table~\ref{tab:gfits} we give the average of the sizes obtained with these 3
methods; uncertainties are estimated from the scatter in the results.  

For the measured source sizes, the integrated flux densities correspond to peak
brightness temperatures of approximately 500~K.  Uncertainties in the
integrated fluxes are dominated by the absolute calibration uncertainties of
$\pm~10$\% for Band7 and $\pm$~15\% for Band~9, as given by the ALMA Cycle 1
Technical Handbook, and of $\pm$~15\% estimated for the 230~GHz CARMA data.

The absolute positions of SrcI at 349 and 661~GHz differ by ($-18,+34$) and
($-52,+22$) milliarcseconds from SrcI's expected position in 2014 July, $05^{\rm
h}35^{\rm m}14\fs5164$, $-5\degr22\arcmin30\farcs598$, predicted from the
43~GHz position and proper motions given by \citet{Goddi2011}.  Since the ALMA
data are snapshot observations obtained with a single phase calibrator, not
optimized for astrometry, we consider this to be reasonable agreement.

The 229 GHz flux density in Table~\ref{tab:gfits}, $330\pm50$~mJy, is larger
than the value $310\pm45$~mJy that we reported previously \citep[][
Table~1]{Plambeck2013}; the new value is derived from combined CARMA A and B
array data, whereas the old value was based on A array data alone.  

The spectral energy distribution of SrcI from cm to submm wavelengths will be
discussed later, in Section~\ref{sec:SED}. 

%============================= Fig B7spect ============================
\begin{figure*}[p]
\centering
% trim left bottom right top
\includegraphics[width=0.9\textwidth, clip, trim=1cm 1.5cm 1cm 2.5cm] {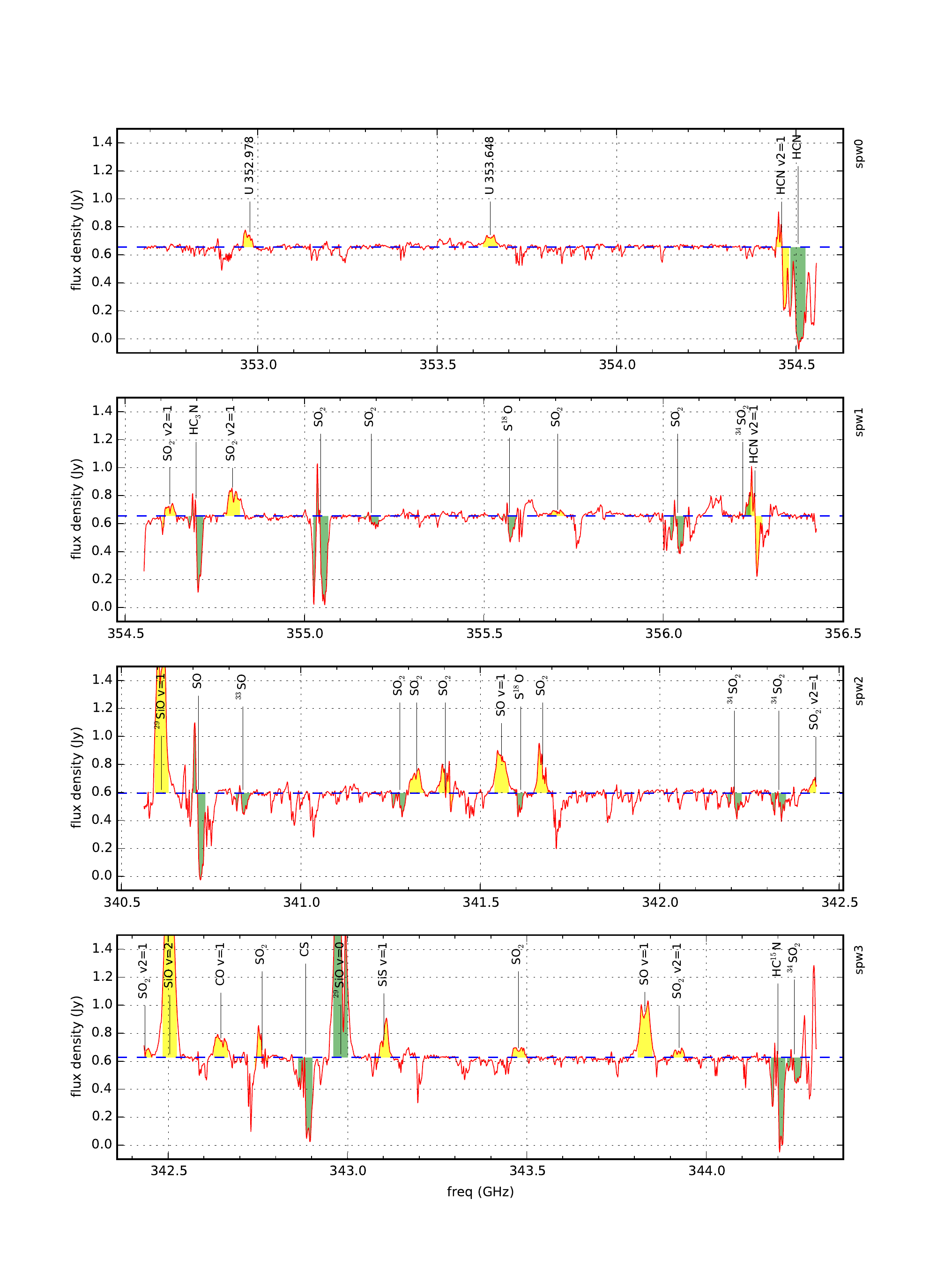} 

\caption{Spectra of SrcI in the four Band 7 spectral windows, generated from
channel maps with an 0\farcs28$\times$0\farcs24 synthesized beam.  The spectral
window is labeled outside the upper right corner of each panel.  The frequency
scale is in the frame of the source, at \Vlsr=5~\kms.  The velocity resolution
is 1.65~\kms.  Flux densities are integrated over an 0\farcs2$\times$0\farcs2
box centered on SrcI.  Blue horizontal dashed lines indicate the continuum
levels.  Selected spectral lines are annotated; their frequencies and upper
state energy levels E$_{\rm U}$ are given in Table~\ref{tab:b7lines}.  Each
line is shaded over a velocity interval of 36~\kms\ centered on the line
frequency; yellow shading indicates E$_{\rm U} > 500$~K; green, E$_{\rm U} <
500$~K.}

\label{fig:B7spect} 
\end{figure*}
%============================= Fig B7spect ============================

%============================= Fig B9spect ============================
\begin{figure*}[p] \centering
% trim left bottom right top
\includegraphics[width=0.9\textwidth, clip, trim=1cm 1.5cm 1cm 2.5cm] {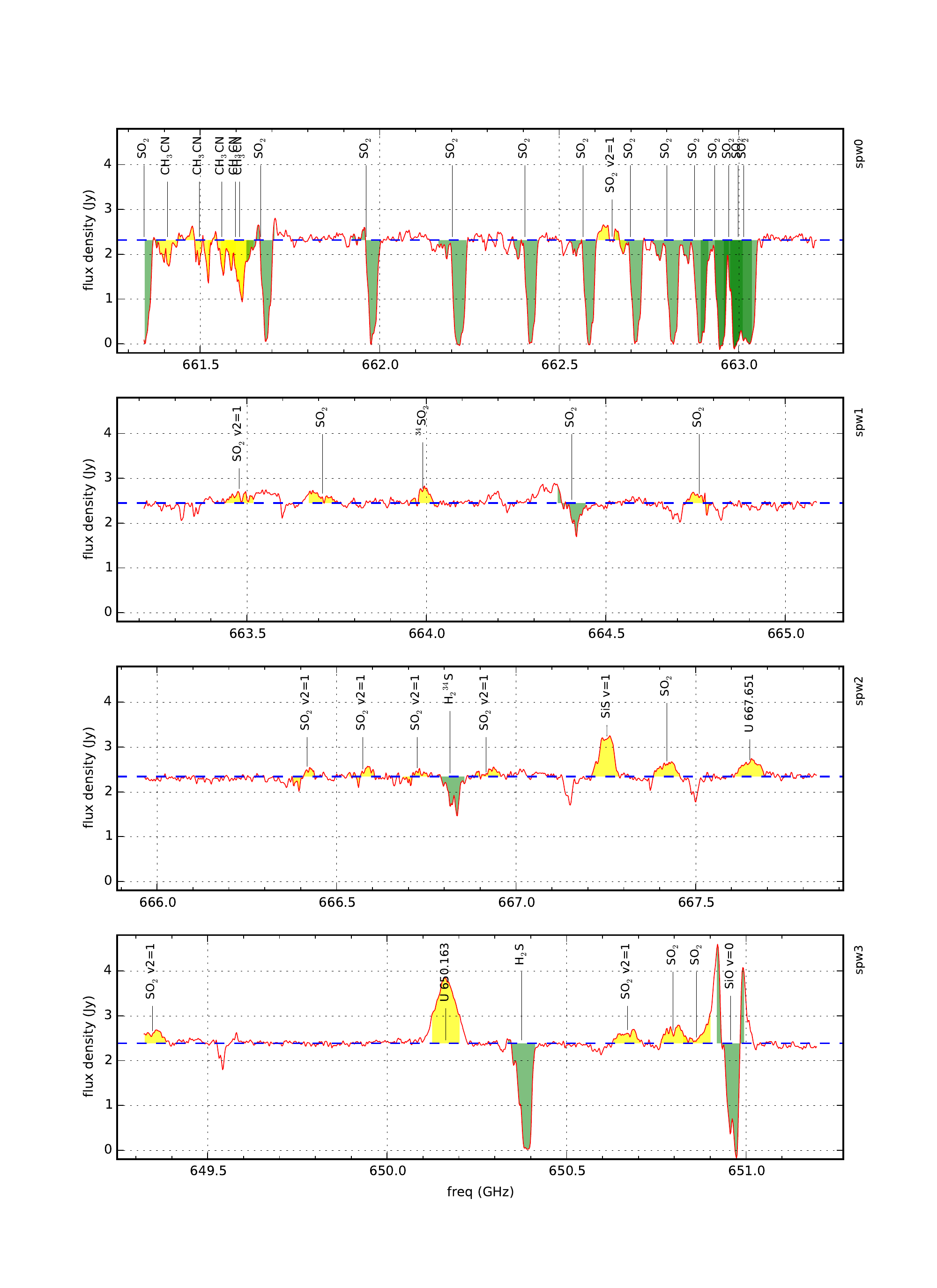} 

\caption{Same as Figure~\ref{fig:B7spect} for the four Band 9 spectral windows.
For these maps the synthesized beam was 0\farcs18$\times$0\farcs14, and the
effective velocity resolution plotted here is 1.8~\kms. Frequencies and upper
state energies for the annotated spectral lines are given in Table~\ref{tab:b9lines}.}

\label{fig:B9spect} 
\end{figure*}
%============================= Fig B9spect ============================

\section{SPECTRAL LINES}
\label{sec:spectral}

Spectral channel images were generated from the self-calibrated visibility data
using baselines $>100$~k$\lambda$ and {\tt robust=0} weighting.  Including
data from baselines shorter than 250~\klambda\ slightly degrades the angular
resolution relative to the continuum images, but improves the signal to noise
ratio. Figures~\ref{fig:B7spect} and \ref{fig:B9spect} display the spectra
integrated over a 0\farcs2$\times$0\farcs2 box centered on SrcI.  These spectra
have been Hanning-smoothed to an effective velocity resolution of 1.7~\kms.
The frequency axes are computed for a source velocity \Vlsr=5~\kms.  This is
the velocity of SrcI determined from previous observations of SiO masers and
H$_2$O lines \citep[e.g.,][]{Wright1995,Hirota2016}.  The blue horizontal
dashed line in each panel indicates the continuum flux density obtained from an
average of the line-free channels in that spectral window.  These flux
densities are lower than those listed in Table~\ref{tab:gfits} because the
0\farcs2$\times$0\farcs2 box does not completely cover the source.  

%============================= Table b7lines ============================
\begin{deluxetable}{rlr} 
\tablewidth{0.65\columnwidth}
\tabletypesize{\small}
\tablecaption{Band 7 Spectral lines}
\tablehead{ 
 \colhead{freq}  & \colhead{molecule} & \colhead{E$_{\rm U}$} \\ 
 \colhead{(GHz)} &  & \colhead{(K)}  
}
\startdata
  340.6119 & $^{29}$SiO v=1 &  1832 \\
  340.7142 & SO &    81 \\
  340.8387 & $^{33}$SO &    87 \\
  341.2755 & SO$_2$ &   369 \\
  341.3233 & SO$_2$ &  1412 \\
  341.4031 & SO$_2$ &   808 \\
  341.5594 & SO v=1 &  1686 \\
  341.6133 & S$^{18}$O &   156 \\
  341.6740 & SO$_2$ &   679 \\
  342.2089 & $^{34}$SO$_2$ &    35 \\
  342.3320 & $^{34}$SO$_2$ &   110 \\
  342.4359 & SO$_2$ v2=1 &  1041 \\
  342.5044 & SiO v=2 &  3595 \\
  342.6476 & CO v=1 &  3117 \\
  342.7616 & SO$_2$ &   582 \\
  342.8829 & CS &    66 \\
  342.9808 & $^{29}$SiO v=0 &    74 \\
  343.1010 & SiS v=1 &  1236 \\
  343.4767 & SO$_2$ &  2068 \\
  343.8294 & SO v=1 &  1677 \\
  343.9237 & SO$_2$ v2=1 &  1058 \\
  344.2001 & HC$^{15}$N &    41 \\
  344.2453 & $^{34}$SO$_2$ &    89 \\
  352.9780 & {\bf unknown} \\
  353.6480 & {\bf unknown} \\
  354.4604 & HCN v2=1 &  1067 \\
  354.5055 & HCN &    43 \\
  354.6242 & SO$_2$ v2=1 &  1852 \\
  354.6975 & HC$_3$N &   341 \\
  354.8000 & SO$_2$ v2=1 &   928 \\
  355.0455 & SO$_2$ &   111 \\
  355.1865 & SO$_2$ &   180 \\
  355.5711 & S$^{18}$O &    93 \\
  355.7055 & SO$_2$ &  1143 \\
  356.0406 & SO$_2$ &   230 \\
  356.2222 & $^{34}$SO$_2$ &   320 \\
  356.2556 & HCN v2=1 &  1067 
\enddata 
\label{tab:b7lines}
\end{deluxetable} 
%============================= Table b7lines ============================

%============================= Table b9lines ============================
\begin{deluxetable}{rlr} 
\tablewidth{0.65\columnwidth}
\tabletypesize{\small}
\tablecaption{Band 9 Spectral Lines}
\tablehead{ 
 \colhead{freq}  & \colhead{molecule} & \colhead{E$_{\rm U}$} \\ 
 \colhead{(GHz)} &  & \colhead{(K)}  
}
\startdata
  649.3467 & SO$_2$ v2=1 &  2037 \\
  650.1630 & {\bf unknown} \\
  650.3742 & H$_2$S &   282 \\
  650.6693 & SO$_2$ v2=1 &  1580 \\
  650.7967 & SO$_2$ &  1075 \\
  650.8620 & SO$_2$ &  1090 \\
  650.9561 & SiO v=0 &   250 \\
  661.3436 & SO$_2$ &   333 \\
  661.4088 & CH$_3$CN &   702 \\
  661.4968 & CH$_3$CN &   652 \\
  661.5597 & CH$_3$CN &   616 \\
  661.5975 & CH$_3$CN &   595 \\
  661.6101 & CH$_3$CN &   588 \\
  661.6683 & SO$_2$ &   313 \\
  661.9622 & SO$_2$ &   295 \\
  662.2027 & SO$_2$ &   277 \\
  662.4043 & SO$_2$ &   261 \\
  662.5669 & SO$_2$ &   245 \\
  662.6479 & SO$_2$ v2=1 &  1443 \\
  662.6976 & SO$_2$ &   230 \\
  662.7995 & SO$_2$ &   217 \\
  662.8770 & SO$_2$ &   204 \\
  662.9336 & SO$_2$ &   192 \\
  662.9728 & SO$_2$ &   181 \\
  662.9977 & SO$_2$ &   171 \\
  663.0144 & SO$_2$ &   146 \\
  663.4789 & SO$_2$ v2=1 &  1476 \\
  663.7109 & SO$_2$ &  1496 \\
  663.9904 & $^{34}$SO$_2$ &   772 \\
  664.4047 & SO$_2$ &   217 \\
  664.7604 & SO$_2$ &   921 \\
  666.4179 & SO$_2$ v2=1 &  1411 \\
  666.5734 & SO$_2$ v2=1 &  1380 \\
  666.7238 & SO$_2$ v2=1 &  1284 \\
  666.8162 & H$_2$$^{34}$S &   245 \\
  666.9162 & SO$_2$ v2=1 &  1899 \\
  667.2525 & SiS v=1 &  1680 \\
  667.4198 & SO$_2$ &  1649 \\
  667.6510 & {\bf unknown} 
\enddata 
\label{tab:b9lines}
\end{deluxetable} 
%============================= Table b9lines ============================

A multitude of molecular lines are evident in the spectra.  These include many
narrow absorption features that originate in relatively cool
(T~$\lesssim200$~K) foreground gas that is not in close proximity to SrcI.  It
is beyond the scope of this paper to identify all of these lines.  The velocity
of the absorbing gas is uncertain by a few \kms, so often a feature matches 5
or 10 candidate transitions in the Splatalogue spectral line
database\footnote{www.spatalogue.net}.  Only a global fit to the entire
spectrum, to confirm that {\it all} expected transitions of a given molecule
are present, can winnow out the most likely identifications. 

Here we are primarily interested in broad ($\Delta$v $\gtrsim30$~\kms) spectral
features that originate from SrcI or its outflow.  Many of these lines are
identified in Figures~\ref{fig:B7spect} and \ref{fig:B9spect}; their rest
frequencies and upper state energy levels are given in
Tables~\ref{tab:b7lines}~and~\ref{tab:b9lines}.  \mbox{Sulfur-} and
\mbox{silicon-rich} molecules such as SiO, SO, SO$_2$, SiS, and H$_2$S are
particularly prominent.  A few lines do not match any plausible transitions in
the Splatalogue database; they are listed in boldface in
Tables~\ref{tab:b7lines}~and~\ref{tab:b9lines}.

In the figures, transitions with upper state energies E$_{\rm U}<500$~K are shaded in
green; these appear mainly in absorption against the SrcI continuum. Transitions 
with E$_{\rm U}> 500$~K are shaded in yellow; these appear mainly in emission.
The shaded widths are 36~\kms\ in all cases, equal to the full width of the SiO
or H$_2$O maser emission from SrcI \citep[e.g.,][]{Genzel1981,Wright1995}.  

% ============================= Fig recomb3 ============================
\begin{figure} \centering
% trim left bottom right top
\includegraphics[width=1.0\columnwidth, clip, trim=1.5cm 15.8cm 12.5cm 2.5cm] {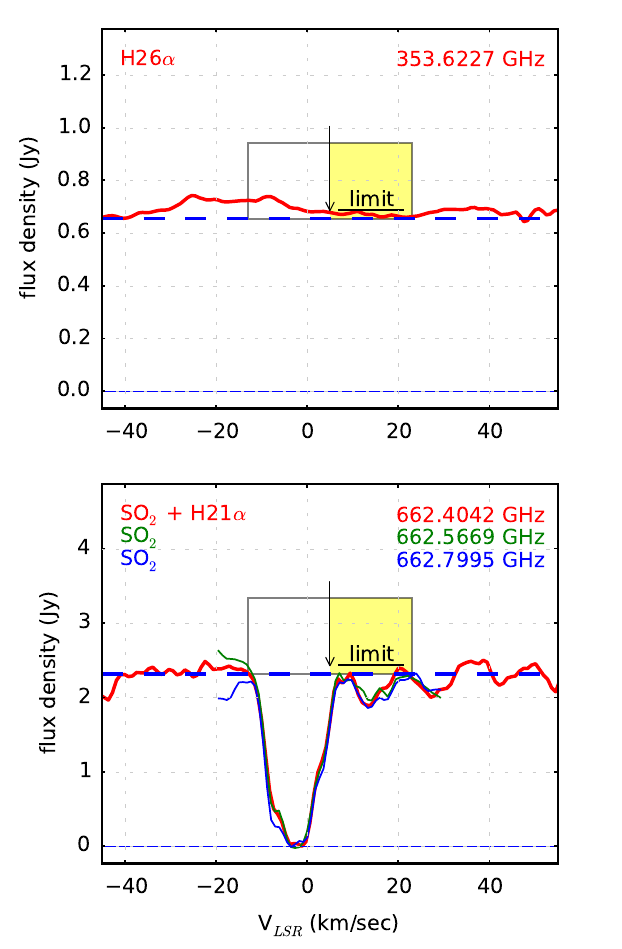} 

\caption{SrcI spectra centered on the H26$\alpha$ and H21$\alpha$ recombination
line frequencies.  As in Figures~\ref{fig:B7spect} and \ref{fig:B9spect}, flux
densities are summed over 0\farcs2$\times$0\farcs2 boxes centered on SrcI.
Black arrows indicate the expected line center velocities; gray rectangles, the
expected linewidths.  In the top panel, an unidentified emission line centered
at -18~\kms\ intrudes into the H26$\alpha$ rectangle.  In the bottom panel, the
17(7,11)-17(6,12) transition of SO$_2$ coincides exactly with the recombination
line frequency, but SO$_2$ absorption obscures only the blueshifted half of the
H21$\alpha$ rectangle; spectra of two neighboring SO$_2$ lines are overplotted
for comparison.  Upper limits on the recombination line intensities are based
on the redshifted velocity ranges indicated by shading.  } 

\label{fig:recomb3} 
\vspace{12pt} 
\end{figure}
% ============================= Fig recomb3 ============================

\subsection{Hydrogen Recombination Lines}  
\label{sec:recomb}

A primary goal of the ALMA observations was to search for emission from the
H26$\alpha$ and H21$\alpha$ recombination lines, at 353.6228 and
662.4042~GHz respectively.  A detection of either transition would confirm the
presence of ionized hydrogen in SrcI, strongly bolstering the case that the
source is a hypercompact HII region.  

Figure~\ref{fig:recomb3} displays 100~\kms\ wide slices of the SrcI spectrum
centered on the recombination line frequencies.  We expect that, if present,
the recombination lines would be centered at \Vlsr\ $\sim 5$~\kms\ with widths
$\geq 36$~\kms, similar to the molecular emission lines.  This velocity
range is indicated by the light gray rectangle in each panel.  There are no
indications of emission peaks at the expected frequencies, although neither
spectrum is perfectly clean---in the upper panel, a weak, unidentified emission
line at 353.648~GHz intrudes into the blueshifted portion of the H26$\alpha$
rectangle, while in the lower panel, strong absorption by the
17(7,11)--17(6,12) transition of SO$_2$ at 662.4043 GHz obscures the blueshifted
half of the H21$\alpha$ rectangle. 

One might expect that it would be impossible to set an upper limit on the
strength of the H21$\alpha$ line, since the interfering SO$_2$ transition
coincides with it almost perfectly in frequency.  However, as discussed in
Section~\ref{sec:absorption} below, SO$_2$ absorption originates primarily in
the outflow from SrcI, so it affects only blueshifted velocities.  For \Vlsr$>
5$~\kms\ the outflowing gas lies primarily {\it behind} SrcI, so it should not
attenuate recombination line emission from the source.  The 17(7,11)-17(6,12)
transition is one of a series of SO$_2$ lines near 662~GHz with similar line
strengths and upper state energy levels.  The spectra of two neighboring lines,
the 16(7,9)-16(6,10) and 14(7,7)-14(6,8) transitions, are overlaid in
Figure~\ref{fig:recomb3} to demonstrate the similarity of the three SO$_2$ line
profiles.  There is no excess emission in the redshifted wing of the
17(7,11)-17(6,12) line that that could be attributed to H21$\alpha$.
 
We use the redshifted velocity range $5 <$~\Vlsr$~< 23$~\kms\ (shaded in
Figure~\ref{fig:recomb3}) to establish upper limits on the intensities of the
recombination lines.  As shown by the horizontal black lines in each spectrum,
all channels within this velocity range have line + continuum flux
densities that are $<~1.05~\times$ the mean continuum level. This limit on the
line-to-continuum ratio will be used in Section~\ref{sec:hypercompact} to argue
that free-free emission is not a viable explanation for SrcI's continuum.

%============================= Fig outflow ============================
\begin{figure} \centering
% trim left bottom right top
\includegraphics[width=0.95\columnwidth, clip, trim=6cm 2cm 3cm 1.7cm] {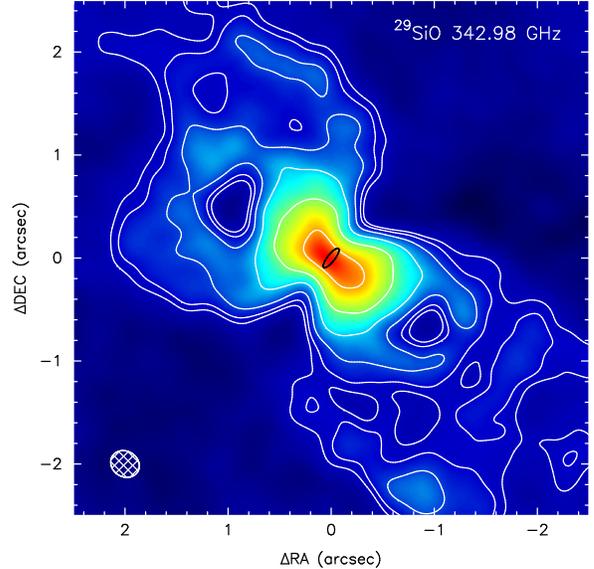} 

\caption{Map of the v=0, J=8-7 line of $^{29}$SiO over the velocity range
$0<$\Vlsr$<10$~\kms, showing the bipolar outflow from SrcI.  The synthesized
beam, shown in the lower left corner, is 0\farcs29$\times$0\farcs25 at
PA~58\degr.  The contours are 0.05, 0.1, 0.2, 0.4, 0.8, 1.6 Jy/beam (7.3 to
230~K).  The black,  0\farcs23 $\times$ 0\farcs07 ellipse at the center shows
the size and location of the {\it model} continuum disk.}

\label{fig:outflow} 
\end{figure}
%============================= Fig outflow ============================

\subsection{Bipolar Outflow}
\label{sec:outflow}

Previous observations
\citep[e.g.,][]{Plambeck2009,Niederhofer2012,Greenhill2013} have established
that the bipolar outflow from SrcI is most clearly visible in images of SiO
lines in the v=0 vibrational state.  Our data include two such transitions, the
J=8-7 $^{29}$SiO line at 342.981~GHz and the J=15-14 SiO line at 650.956~GHz.
The lower frequency, J=8-7, data contain more short antenna spacings (see
Figure~\ref{fig:uvcov}), so do a better job of sampling the large scale
structure of the outflow.  An image of this transition integrated over the
velocity range $0<$\Vlsr$<10$~\kms\ is shown in Figure~\ref{fig:outflow}.  All
the  visibility data, down to the shortest, 35~\klambda\ baselines, were used
to make this image.  No other lines in our data show this extended bipolar
pattern.  

% ============================= Fig absorption ============================
\begin{figure} \centering
% trim left bottom right top
\includegraphics[width=1.0\columnwidth, clip, trim=1.5cm 17cm 10cm 2.5cm] {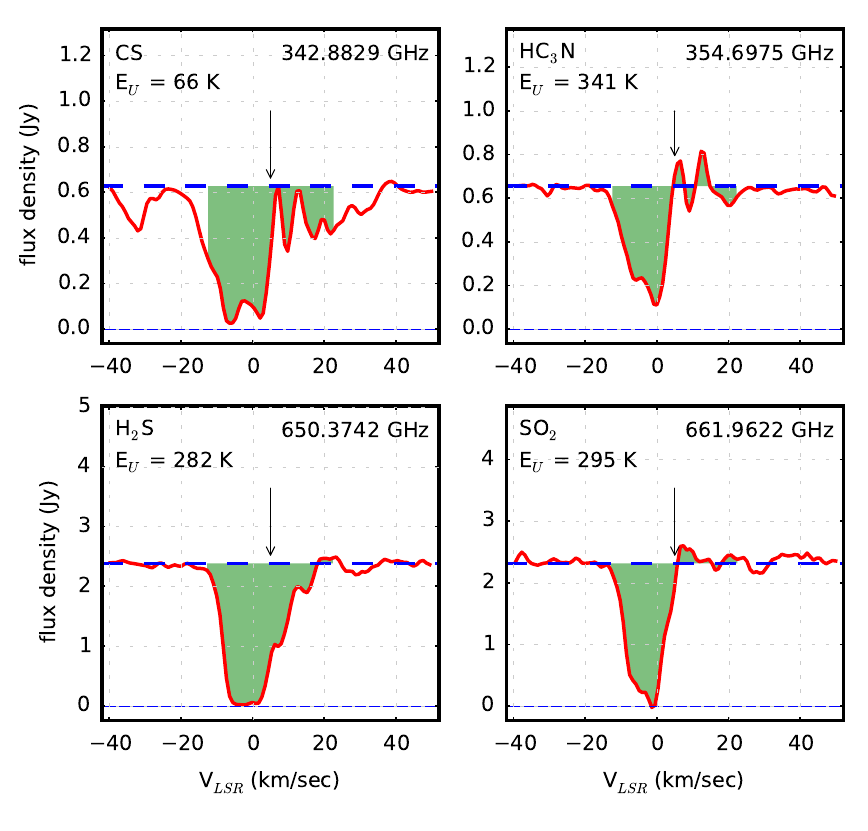} 

\caption{Examples of molecular lines exhibiting blueshifted absorption by the
SrcI outflow.  Line frequencies and upper energy levels are given.  As in
Figures~\ref{fig:B7spect} and \ref{fig:B9spect}, flux densities are summed over
0\farcs2$\times$0\farcs2 boxes centered on SrcI, and the dashed blue lines
indicate the continuum levels.  Arrows indicate the LSR velocity of SrcI,
5~\kms; 36~\kms\ wide ranges centered on this velocity are shaded.  }

\label{fig:absorption} 
\vspace{12pt} 
\end{figure}
% ============================= Fig absorption ============================

\subsection{Absorption Lines}
\label{sec:absorption}

Almost all of the lines with upper state energies E$_{\rm U} < 500$~K appear in
absorption against the SrcI continuum.  The absorption is highly asymmetric.
It is limited primarily to {\it blueshifted} ($-13 <$ \Vlsr $< 5$~\kms)
velocities; i.e., it originates in gas that is moving toward us from SrcI.
Figure~\ref{fig:absorption} shows several examples.  Since the velocity width
of the absorption is comparable with the halfwidth of the SiO v=0 emission, we
associate it with material within the SrcI outflow, rather than from unrelated
foreground gas along the line of sight.

Although absorption is apparent in lines of many abundant molecules
like HCN, HC$^{15}$N, and HC$_3$N, it is especially prominent in transitions of
sulfur-bearing species such as SO, SO$_2$, H$_2$S, and their isotopologues.

In 1\farcs5 resolution ALMA Band 6 science verification data, \citet{Wu2014}
found evidence for {\it redshifted} CH$_3$OH absorption toward SrcI, indicative
of infall.  Presumably this infall occurs in the outer layers of the
Kleinmann-Low Nebula, on angular scales $\gtrsim\,$3\arcsec\ that are poorly 
sampled by our observations.

\subsection{Emission Lines}
\label{sec:emission}

Lines with upper state energy levels E$_{\rm U} > 500$~K, shaded in yellow in
Figures~\ref{fig:B7spect} and \ref{fig:B9spect}, almost always are seen in
emission.  Generally these lines are symmetric about \Vlsr=5~\kms, with full
widths of 30-40~\kms.  Most of them can be identified with transitions of the
molecules SO, SO$_2$, SiS, and SiO, often in excited vibrational states.  We
believe that the line identifications are correct because essentially all of the
stronger transitions of these molecules that are listed in the Splatalogue
database correspond to emission features in the spectra.
For example, within the frequency ranges covered by our observations, the
Splatalogue lists 10 SO$_2$ transitions in the ground vibrational state with
intensities $> 10^{-5}$ (at 300 K) and upper state energies between 500 and
2500~K. Of these, 8 are detected.  The undetected transitions are at 650.862
GHz, buried under the wing of the bright SiO v=0 line, and at 661.511 GHz,
amidst a jumble of CH$_3$CN absorption features.  Similarly, we detect 12 out
of 16 strong SO$_2$ transitions in the v2=1 vibrational level. 

Of the $>30$ identified emission lines, all but two have upper state energy
levels between 500 and 2100~K.  The two exceptions are the v=1, J=3-2 CO line
at 342.648~GHz (E$_{\rm U}$ = 3117~K) and the v=2, J=8-7 SiO line at
342.504~GHz (E$_{\rm U}$ = 3595~K).  These upper state energies are comparable
to those of the vibrationally excited H$_2$O lines at 232.687~GHz (E$_{\rm U}$
= 3462~K) and 336.228~GHz (E$_{\rm U}$ = 2955~K) previously identified by
\citet{Hirota2012,Hirota2014} in SrcI.

\begin{figure} \centering
% trim left bottom right top
\includegraphics[width=1.0\columnwidth, clip, trim=1.5cm 2cm 7.2cm 2.5cm] {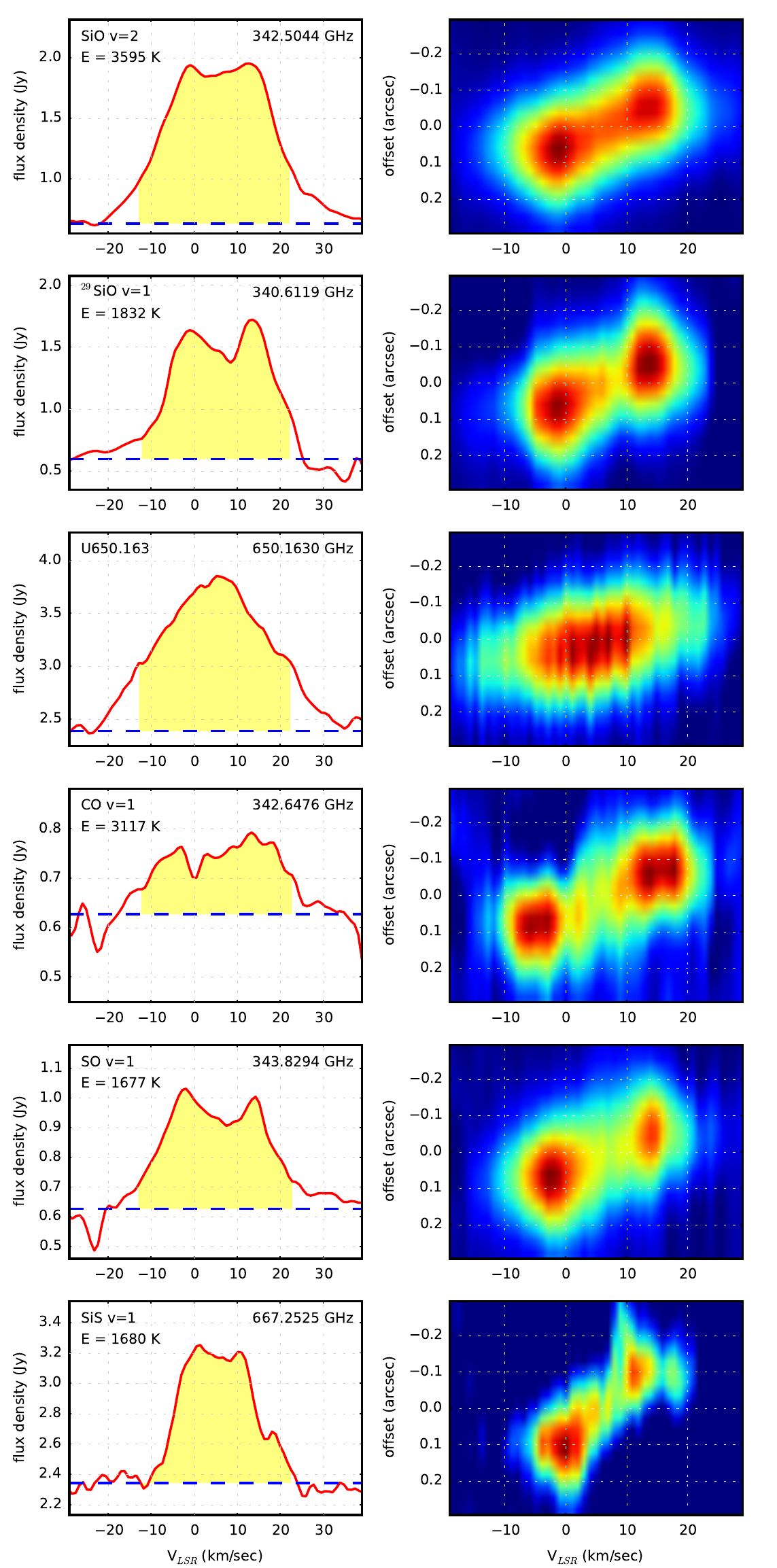} 

\caption{Spectra and position-velocity plots for selected emission lines.  As
in previous figures, spectra are integrated over an 0\farcs2 box centered on
SrcI, and are shaded over a 36~\kms\ range centered at \Vlsr=5~\kms.  
Position-velocity diagrams are measured along the major axis of the SrcI disk,
at PA 142\degr; negative offsets are to the NW, positive offsets to the SE.
The continuum was subtracted from the spectral line channel images before
generating the position-velocity plots.  }

\label{fig:snip_pv} 
\end{figure}

Four of the emission lines are unidentified.  The strongest of these is a broad
(fullwidth 50~\kms), centrally peaked feature at 650.163 GHz.  As shown in
Figure~\ref{fig:snip_pv}, the position-velocity diagram of this line is
similar to those of the SiO v=1 and v=2 lines, hinting that it may originate
from a vibrationally excited transition with $E_{\rm U} > 2500$~K. 

%============================= Fig snip_pv ============================
%============================= Fig snip_pv ============================

%============================= Fig centroids ============================
\begin{figure} \centering
% trim left bottom right top
\includegraphics[width=1.0\columnwidth, clip, trim=0cm 1cm 3.5cm 1.5cm] {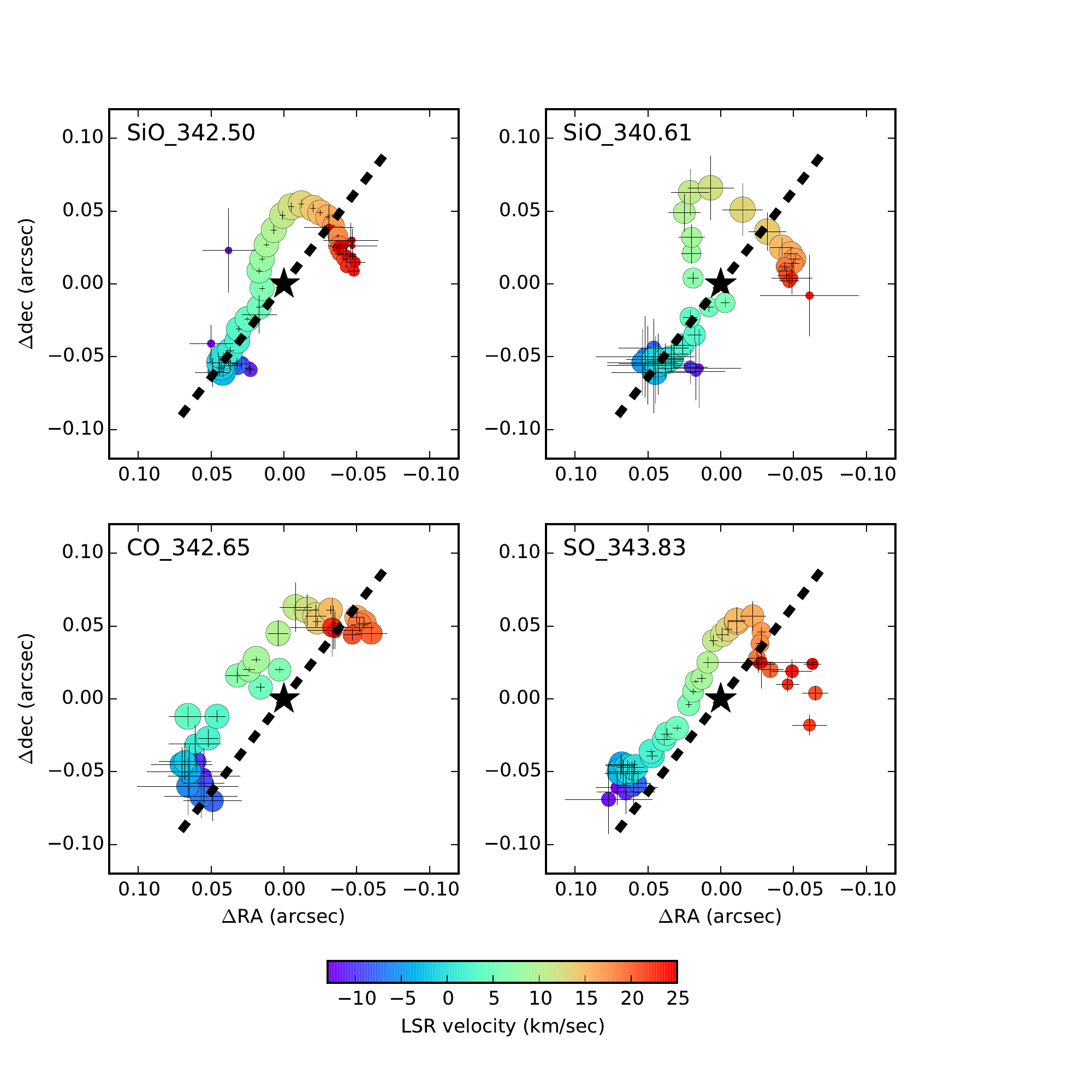} 

\caption{Velocity centroid maps for selected spectral lines.  The area of each
circle is proportional to the flux in that channel; colors indicate the radial
velocity.   The 0\farcs23 long dotted line in each panel represents the
midplane of the continuum disk.  }

\label{fig:centroids} 
\end{figure}
%============================= Fig centroids ============================

\subsection{Central Mass Estimate}
\label{sec:mass}

Figure~\ref{fig:snip_pv} presents spectra of 6 of the brightest emission lines,
and the corresponding position-velocity diagrams measured along the major axis
of the continuum disk, at PA~142\degr.  All the lines show a velocity gradient
along the disk.  Except for U650.163, the spectra are double-peaked, and the
brightest emission is offset by 0.05-0.10\arcsec\ from the origin.
The double-peaked line profiles and the absence of high velocity gas toward the
central position suggest that molecular line emission originates from a torus
or ring, rather than a filled disk.

Figure~\ref{fig:centroids} plots the position centroids of individual 1~\kms\
wide spectral channels for 4 selected molecular lines.  The offsets were
derived from 2-dimensional Gaussian fits to 0\farcs26 resolution images.
Centroids are plotted only for channels where the fits have formal
uncertainties smaller than 0\farcs03.  The 0\farcs23-long dotted line at PA
142\degr\ in each panel represents the midplane of the continuum disk.  These
plots resemble the centroid map of the 336~GHz transition of vibrationally
excited H$_2$O \citep{Hirota2014}, and are consistent with the much higher
resolution SiO maser spot maps \citep{Kim2008,Matthews2010}.  The highest
velocity gas tends to be localized along the edges of the bipolar outflow.
 
%============================= Fig rotcurvefit ============================
\begin{figure} \centering
% trim left bottom right top
\includegraphics[width=0.9\columnwidth, clip, trim=1cm 1cm 1cm 2cm] {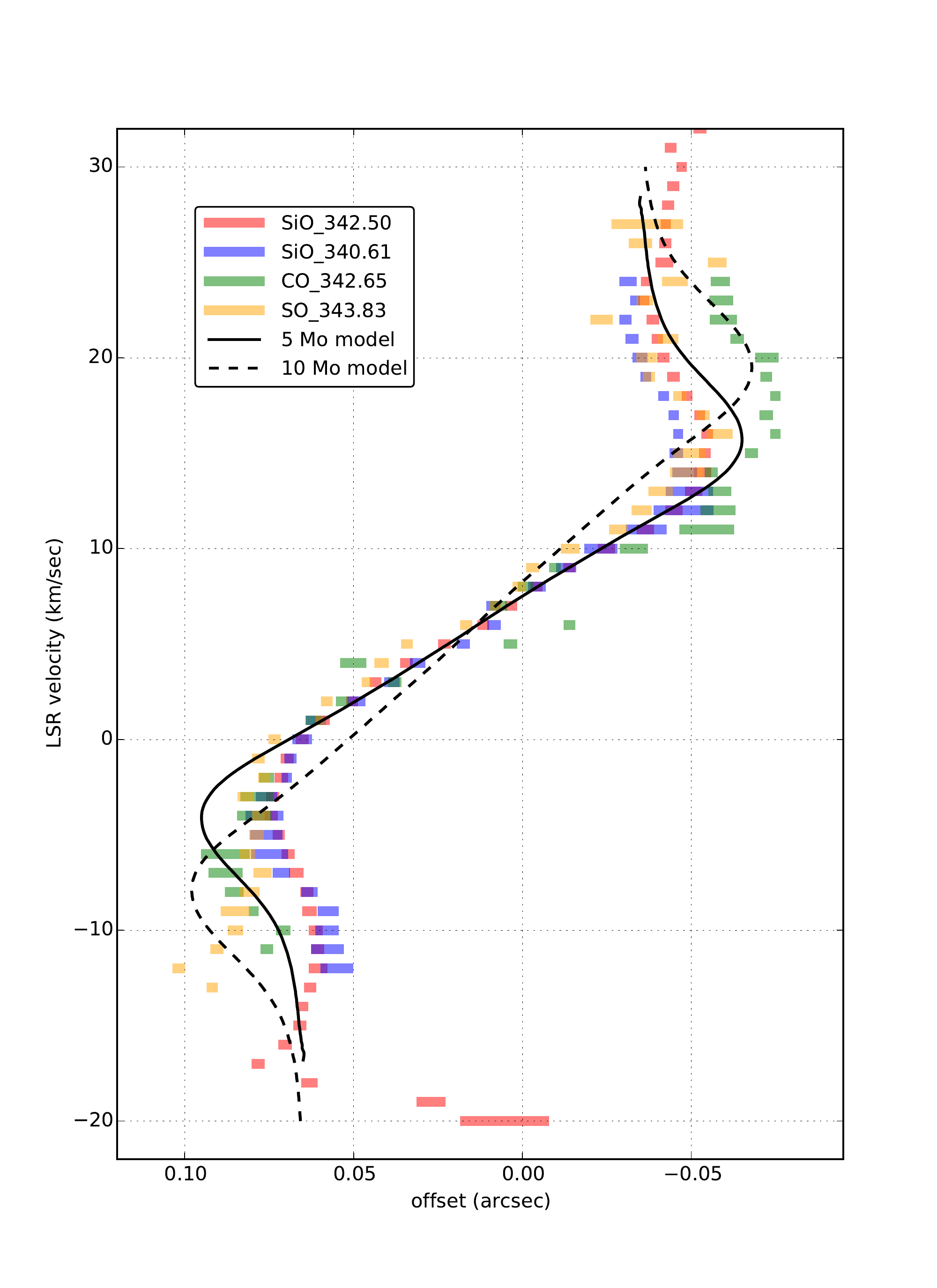} 

\caption{Position-velocity diagrams along the disk midplane for the spectral lines shown
in Figure~\ref{fig:centroids}.  Horizontal bars indicate the $\pm 1\sigma$
formal errors in the position offsets. 
Model rotation curves for 5~\Msol\ and 10~\Msol\ central objects are shown by
solid and dashed lines.  The model assumes optically thin emission from gas in
a rotating, edge-on ring with 20~AU inner radius, 50~AU outer radius, and
4~\kms\ turbulent linewidth.}

\label{fig:rotcurvefit} 
\end{figure}
%============================= Fig rotcurvefit ============================

%============================= Fig mc3 ============================
\begin{figure*} \centering
% trim left bottom right top
\includegraphics[width=0.9\textwidth, clip, trim=1cm 8.5cm 1cm 2cm] {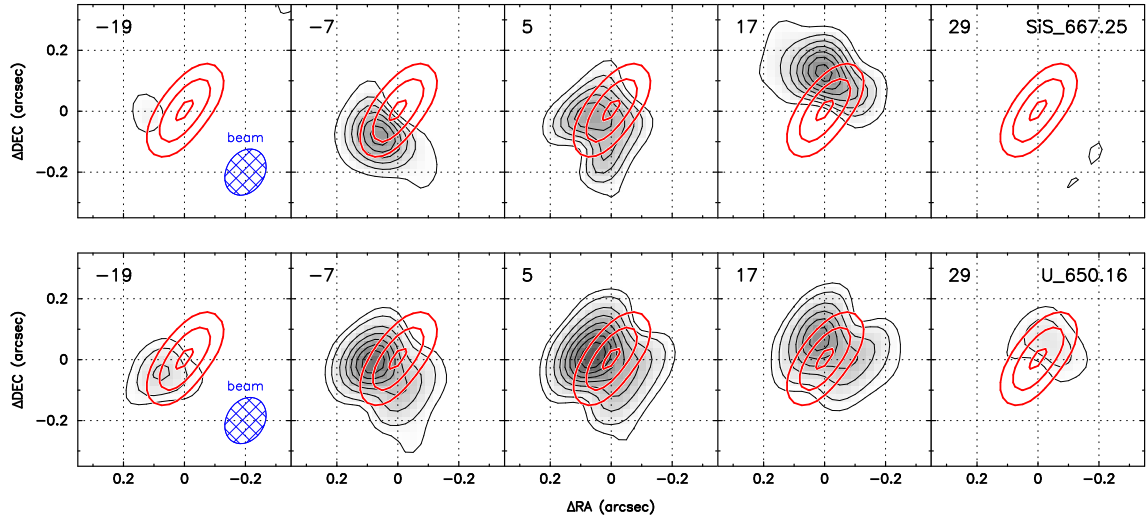} 

\caption{Comparison of spectral line (halftone images, thin black contours) and
661 GHz continuum emission (thick red contours) toward SrcI.  Top panels show
the SiS transition at 667.252~GHz (contour levels 0.04, 0.08, $\ldots$, 0.28
Jy/beam); bottom panels show the unidentified spectral line at 650.163 GHz
(contour intervals 0.05, 0.1, 0.2, $\ldots$, 0.8 Jy/beam).  Each channel map
covers a 12~\kms\ wide interval centered at the LSR velocity indicated in the
upper left corner.  The continuum contours are 0.5, 1.0, 1.5 Jy/beam.  The
continuum has been subtracted from each spectral channel. The FWHM of the
synthesized beam, 0\farcs16$\times$0\farcs12, is shown by the hatched blue ellipse
in the lefthand panels.  Spectral line emission extends above and below the
continuum disk.}

\label{fig:mc3} 
\end{figure*}
%============================= Fig mc3 ============================

We projected the channel centroids in Figure~\ref{fig:centroids} onto the disk
midplane in order to generate the rotation curves shown in
Figure~\ref{fig:rotcurvefit}.  To estimate the mass of the central object, we
simulated these rotation curves using a model similar to the one used by
\citet{Hirota2014}.  The model assumes optically thin emission from gas in
Keplerian rotation around a central point mass.  The gas has uniform
temperature and density, and is confined to a ring that is viewed exactly
edge-on.  We fixed the outer radius of the ring to 50~AU to match the radius of
the continuum disk, but experimented with different inner radii.  A turbulent
linewidth of 4~\kms\ was assumed.  For each velocity we numerically integrated
the emission from all cells in the ring in order to compute the offset of the
emission centroid from the origin.

Note that, although the spectral line profiles are symmetric around
\Vlsr$\sim$5~\kms, the blueshifted channel centroids typically lie farther from
the origin than the redshifted centroids, so that the dynamical center of
rotation used by the model is 0\farcs015 (6 AU) from the center of the
continuum source.  Higher angular resolution is required to see if this offset
is real.

We modeled the rotation curves for a range of central masses and inner radii.
Central masses of 5-7~\Msol\ matched the data best, particularly for offsets
$<$0\farcs05 where the rotation curves of all 4 spectral lines are consistent.
Figure~\ref{fig:rotcurvefit} shows the model results for 5~\Msol\ and
10~\Msol\ central objects and an inner ring radius of 20~AU. The 10~\Msol\
simulation (dashed) is steeper than the observed rotation curve at small radii.
This is similar to the result obtained by \citet{Hirota2014} for the 336~GHz
H$_2$O line in SrcI.  

Of course the model is highly idealized, particularly since it utilizes
centroids to represent emission regions that are unresolved by the synthesized
beam.  However, an estimated central mass of 5-7~\Msol\ is consistent with that
derived by \citet{Matthews2010} from high resolution SiO maser maps.  The
central mass could be larger if the plane of the disk is inclined with respect
to the line of sight, but the observed axial ratio of the continuum source,
\mbox{0\farcs23 : 0\farcs07}, implies that the disk is tilted by no more than
about 20\degr, which changes the central mass by less than 10 percent.

\subsection{Outflow or Disk Midplane Emission?}
\label{sec:source}

\citet{Hirota2014} inferred that the 336~GHz H$_2$O line was emitted from hot
gas in the midplane of the SrcI disk because the velocity channel centroids for
this line were tightly clustered along the major axis of the continuum disk.
Their observations were made with an 0\farcs4 beam, however.  In contrast, VLBI
observations with 0.5 millarcsecond resolution show that 43~GHz SiO masers
avoid a 0.05\arcsec\ wide band along the disk midplane
\citep{Kim2008,Matthews2010}.

Although our images have inadequate resolution to say whether line emission
avoids the disk midplane, they do clearly show that line emission is spatially
extended relative to the continuum.  This is evident in Figure~\ref{fig:mc3},
which overlays Band 9 line channel maps and continuum contours.  These images
were generated from visibilities measured on baselines longer than
300~\klambda\ in order to obtain 0\farcs13 resolution along the minor axis of
the disk.  The continuum has been subtracted from each line channel.  The
velocity gradient along the major axis of the disk is clearly apparent.
Because line emission extends outward from the top and bottom surfaces of the
disk, we associate it with the base of the bipolar outflow, or with the
atmosphere of the disk.

In Sections~\ref{sec:dust} and \ref{sec:model} we argue that the continuum
radiation from the disk is mostly thermal emission from warm dust, and
hypothesize that SiO masers originate on the surface of the disk where dust
grains are destroyed.  Thus, it's possible that emission from \mbox{silicon-}
and sulfur-rich molecules like SiS, SO, and SO$_2$ is limited primarily to the
disk surface, while the H$_2$O line mapped by \citet{Hirota2014} originates
mostly from the disk interior.

% ============================= Table rotdiag ============================
\begin{deluxetable}{rrrr} 
\tablewidth{0.75\columnwidth}
\tablecaption{SO$_2$ Integrated Intensities}
\tablehead{ 
 \colhead{E$_{\rm U}$} & \colhead{frequency}  & \colhead{S$_{\rm ij}\mu^2$} & \colhead{$\int T_R dv$}\tablenotemark{a} \\ 
 \colhead{(K)}         & \colhead{(GHz)} & \colhead{(D$^2$)}           & \colhead{(K-km/sec)} 
}
\startdata
\multicolumn {4}{c}{v = 0 vibrational level} \\ [0.05in]
   582 &      342.762 &  50.74 &   221 (6)\phn \\
   679 &      341.674 &  68.31 &   366 (6)\phn \\
   808 &      341.403 &  71.97 &   178 (6)\phn \\
   921 &      664.760 &  31.35 &   107 (20) \\
  1075 &      650.797 &  47.15 &   148 (20) \\
  1143 &      355.705 &  14.61 &    71 (6)\phn \\
  1649 &      667.420 &  89.22 &    30 (20) \\
  2068 &      343.477 &  21.36 &    83 (6)\phn \\
\tableline \\ 
\multicolumn {4}{c}{v2 = 1 vibrational level} \\ [0.05in]
%   928 &      354.800 &  23.77 &   261 \phn(6)\tablenotemark{b} \\
   928 &      354.800 &  23.77 &   261 \phn(6) \\
  1041 &      342.436 &  29.58 &    79 \phn(6) \\
  1058 &      343.924 &  19.56 &    49 \phn(6) \\
  1443 &      662.648 &  52.58 &    52 (20) \\
  1476 &      663.479 &  54.57 &    91 (20) \\
  1580 &      650.669 &  60.99 &    87 (20) \\
  1852 &      354.624 & 102.36 &    93 \phn(6) \\
  2037 &      649.347 &  85.49 &    61 (16) 
\enddata 

\tablenotetext{a}{T$_{\rm R}$ in an 0\farcs2 box centered on SrcI, integrated
over 5$<$\Vlsr$<$20~\kms; uncertainties are in parentheses.
\vspace{6pt}}
\label{tab:rotdiag} 
\end{deluxetable} 
% ============================= Table rotdiag ============================

%============================= Fig rotdiag ============================
\begin{figure}[!h] \centering
% trim left bottom right top
\includegraphics[width=1.0\columnwidth, clip, trim=2cm 1.5cm 3cm 1.5cm] {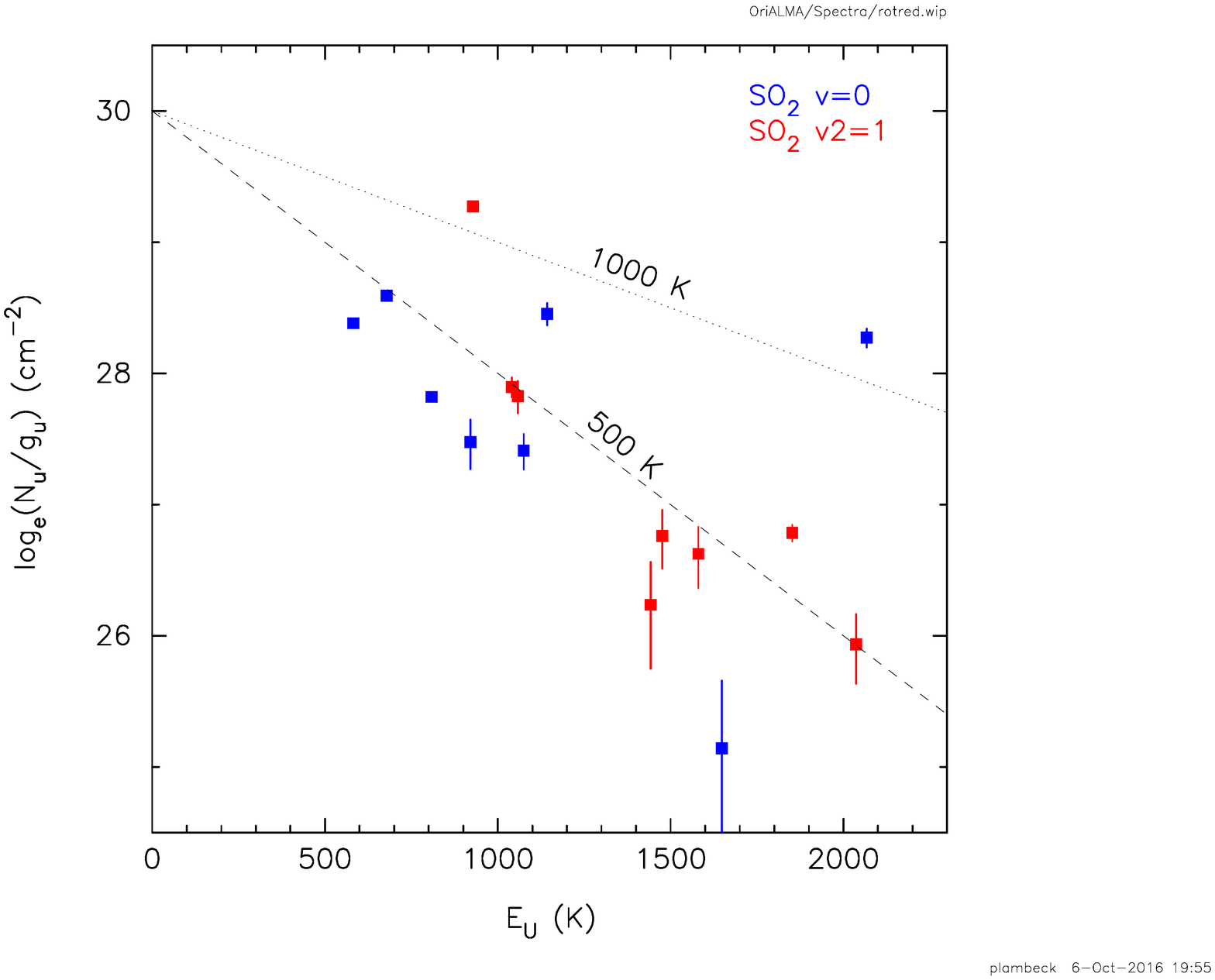} 

\caption{Rotational energy diagram for SO$_2$ emission lines in an 0\farcs2 box
centered on SrcI. Spectral line parameters are given in
Table~\ref{tab:rotdiag}; note that integrated intensities were computed using
only the redshifted velocity range 5-20~\kms. The slope of the dashed line
corresponds to an excitation temperature of 500~K; the dotted line, 1000~K. }

\label{fig:rotdiag} 
\vspace{6pt} 
\end{figure}
%============================= Fig rotdiag ============================

\subsection{Excitation Temperature Analysis}
\label{sec:rotdiag}

The detection of CO, SiO, and H$_2$O lines from vibrationally excited energy
levels indicates that some 3000~K gas exists in SrcI.  However, a rotational
diagram analysis of the SO$_2$ emission lines suggests that the mean excitation
temperature is considerably lower.  Our spectra contain 16 SO$_2$ lines with
E$_{\rm U} >600$~K that appear not to be seriously corrupted by foreground
absorption.  Eight of these are in the ground vibrational state, and eight are
in the v2=1 bending mode vibrational level.  A rotational diagram analysis
\citep[e.g.][]{Blake1987} assumes that the lines are optically thin, that they
originate in the same column of gas, and that the rotational energy ladders are
in local thermodynamic equilibrium.  The SO$_2$ v2=1 level lies only 745~K
above the ground state \citep{Muller2005}, so we will assume that the v=0 and
v2=1 rotational ladders are thermalized at the same temperature.

Table~\ref{tab:rotdiag} lists the upper energy level E$_{\rm U}$, frequency,
line strength $\mu^2S$, and integrated intensity in K-\kms\ for each of the
SO$_2$ lines.  The integrated intensity $W = \int T_R(v) dv$ is computed from
the (continuum-subtracted) radiation temperatures averaged over the velocity
range 5$\,\leq\,$\Vlsr$\,\leq\,$20~\kms, in a 0\farcs2$\times$0\farcs2 box
centered on SrcI; uncertainties are estimated from line-free channels.  We
restrict the analysis to {\it redshifted} velocities because we expect that
this emission originates primarily from the outflow on the far side of SrcI
(see discussion in Section~\ref{sec:absorption}), so we do not have to take
into account molecular absorption of continuum photons from SrcI.

For each line we calculate the quantity $3 k W / (8 \pi^3 \nu S\mu^2) =
N_u/g_u$, the column density per molecular sublevel, where $k$ is Boltzman's
constant and $g_u$ is the degeneracy of the upper energy level.  A plot of
natural log of $N_u/g_u$ against $E_u$ should produce a straight line with
slope $1/T_{\rm ex}$ and intercept $N(SO_2)/Q(T_{ex})$, where $Q(T_{ex})$ is
the rotational partition function at excitation temperature $T_{ex}$.  This
plot is shown in Figure~\ref{fig:rotdiag}.  The slope of the dashed line
corresponds to an excitation temperature of 500~K, and the intercept
corresponds to an SO$_2$ column density  N(SO$_2$) =
$2\times10^{17}$~cm$^{-2}$, where we have taken $Q(500~K) = 15611$, the
combined partition function for the v=0 and v2=1 vibrational states, from the
Cologne Spectral Line Database\footnote{www.astro.uni-koeln.de}.  Although
there is a great deal of scatter in the plot, it appears that little of the gas
is at temperatures greater than about 1000 K (dotted line in
Figure~\ref{fig:rotdiag}).

\section{DISCUSSION}

The observations described above provide important clues to the nature of SrcI.
We first consider the origin of the radio continuum, concluding that it is
primarily thermal dust emission from a disk or torus, optically thick
even at frequencies as low as 43~GHz.  We then compare the measured properties
of SrcI with the predictions of the dynamical decay model.

\subsection{Continuum Spectral Energy Distribution}
\label{sec:SED}

\begin{figure} \centering
% trim left bottom right top
\includegraphics[width=0.9\columnwidth, clip, trim=0.5cm 0.5cm 1cm 1cm] {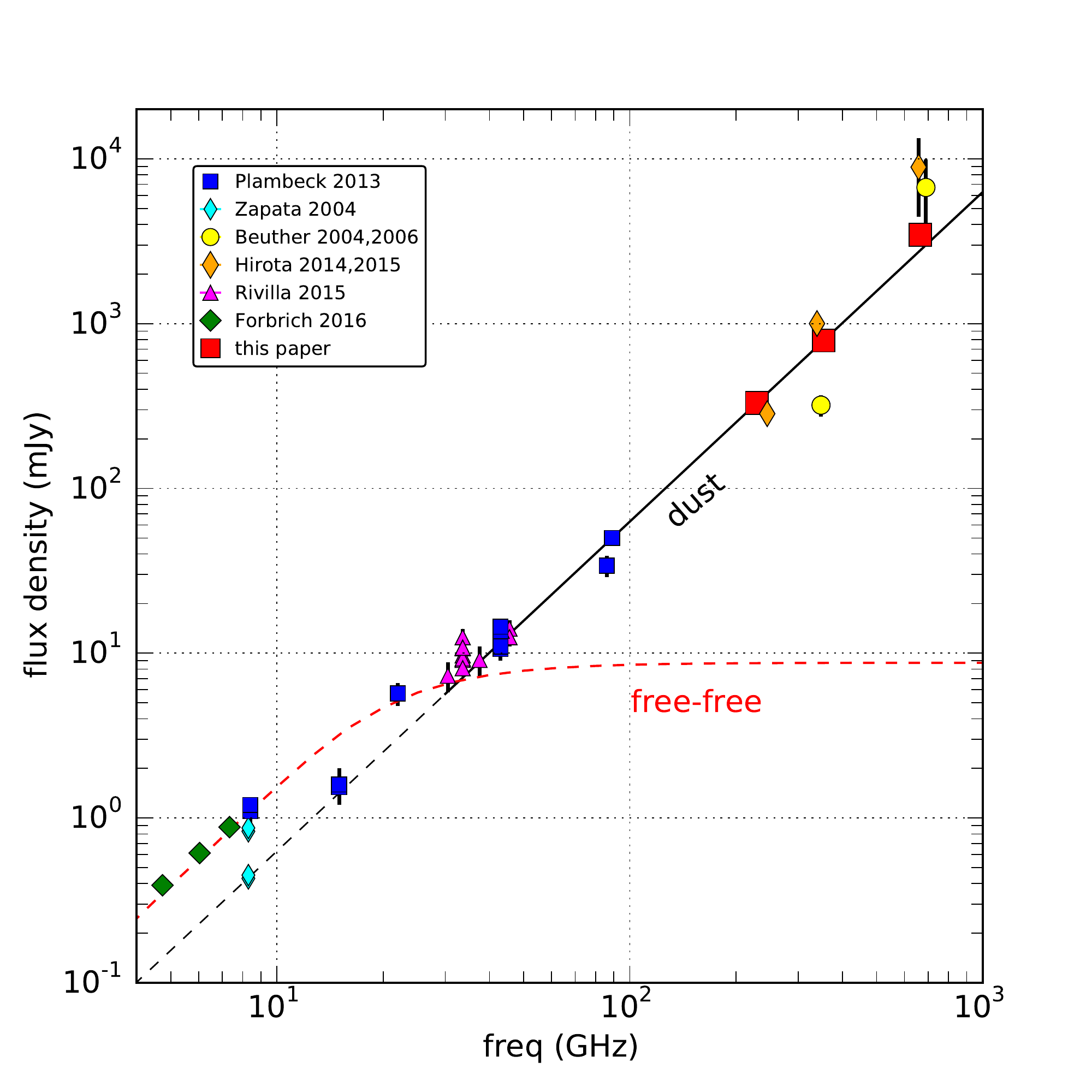} 

\caption{Spectral energy distribution of SrcI from cm to submm wavelengths.
References are given in the legend; red squares indicate the fluxes from this paper.
From 43 to $> 350$~GHz the data follow a $\nu^2$ curve
(black line) that is fit to a flux density of 330~mJy at 229~GHz.  In
Section~\ref{sec:dust} we argue that this emission originates from optically
thick dust.  At lower frequencies free-free emission (red dashed curve)
probably becomes important. }

\label{fig:sed} 
\vspace{6pt} 
\end{figure}

Figure~\ref{fig:sed} plots the continuum flux densities measured for SrcI from
4 to 690~GHz.  The results from this paper are shown as red squares.  The
remaining data are from the compilation in Table 2 in \citet{Plambeck2013},
supplemented by recent VLA \citep{Rivilla2015,Forbrich2016} and ALMA
\citep{Hirota2015,Hirota2016} results.  Error bars are shown for all the
points, but often are smaller than the size of the plot symbols.

The scatter in the flux densities at mm and submm wavelengths ($\nu \gtrsim
100$~GHz) is indicative of the difficulty in making these measurements.  SrcI
is an extended object, not a true point source, and it is embedded in a clumpy
ridge of dust emission.  When observed with a $> 1$\arcsec\ FWHM synthesized
beam it appears merely as a protrusion on the Hot Core dust ridge---see, e.g.,
\citet{Tang2010}, Figure~1, or \citet{Plambeck2013}, Figure~1.  Even with
subarcsecond angular resolution, the images can be corrupted by negative
sidelobes from extended structures that are not fully sampled by the aperture
synthesis observations.  The well-sampled, 0\farcs2 resolution, ALMA and CARMA
data described in this paper provide the most reliable measurements to date at
these wavelengths.

Source variability may also be responsible for some of the scatter in
Figure~\ref{fig:sed}.  \citet{Plambeck2013} presented tentative evidence that
SrcI has brightened relative to BN at 43 and 86 GHz since 1995, and
\citet{Rivilla2015} found indications of variability on time scales of just a
few hours at 34~GHz.

As shown by the solid line in Figure~\ref{fig:sed}, SrcI's flux density scales
approximately as $\nu^2$ from 43~GHz to $>$350~GHz, consistent with emission
from a blackbody that subtends the same solid angle at all frequencies.  The
line shows the flux density that one would receive from a 0\farcs23 $\times$
0\farcs07 FWHM Gaussian source with a peak brightness temperature of 480~K.
Below 43~GHz and above 350~GHz most of the measured flux densities lie above
the $\nu^2$ curve.

% ============================= Fig Flx ==========================
% ============================= Fig Flx ==========================

\subsection{An Optically Thick Hypercompact HII Region?}
\label{sec:hypercompact}

Our data convincingly rule out the possiblity that SrcI is a hypercompact
HII region.  Such an HII region would need to be optically thick even at
660~GHz in order to explain the observed spectral energy distribution and the
absence of a detectable H21$\alpha$ recombination line.  We find, however, that
the Lyman continuum flux required to maintain the ionization of such an HII
region could only be generated by a central star with a luminosity that exceeds
that of the entire Kleinmann-Low Nebula.

To begin, we estimate the minimum free-free optical depth that would be
consistent with the observations.  The recombination line and continuum
opacities are approximated by \citep{Rohlfs2000} 
\begin{eqnarray}
\tau_{\rm L}   & \, = \, & 1.92 \times 10^3\,{T_e}^{-2.5}\,EM\,/\Delta \nu_{\rm kHz} \\ 
\tau_{\rm ff}  & \, = \, & 8.24 \times 10^{-2}\,T_e^{-1.35}\,\nu^{-2.1}\,EM\,a(\nu,T) \label{eqn:tauff}
\end{eqnarray} 
where $T_e$ is the electron temperature in K, $\nu$ is the frequency in GHz,
$EM = \int \! n_e^2\,dz$ is the emission measure in pc~cm$^{-6}$, $\Delta
\nu_{\rm kHz}$ is the recombination linewidth in kHz, and $a(\nu,T)$ is a
correction factor close to unity.  For $\nu=662.4$~GHz, $T_e=8000$~K, and
$\Delta \nu_{\rm kHz} = 1.1 \times 10^5$ ($\Delta v=50$~\kms), these
expressions predict $\tau_{\rm L}\sim5.7\,\tau_{\rm ff}$.  Thus, in the
optically thin limit the H21$\alpha$ recombination line would be 5.7 times
brighter than the free-free continuum.  In Section~\ref{sec:recomb}, however,
we set an upper limit on the H21$\alpha$ intensity of 5\% of the continuum
level, so
\begin{equation}
1 - e^{-(\tau_{\rm L} + \tau_{\rm ff})} < 1.05\,(1 - e^{-\tau_{\rm ff}}). \label{eqn:3}
\end{equation}
For $\tau_{\rm L}/\tau_{\rm ff} = 5.7$, Equation~\ref{eqn:3} implies a
free-free continuum opacity $\tau_{\rm ff} > 3$.

If the path length through SrcI is 100 AU, comparable to its major diameter,
and if $\tau_{\rm ff} > 3$ and $T_e = 8000$~K, then equation~(\ref{eqn:tauff})
implies that the electron density $n_e > 10^8$~cm$^{-3}$.  These numbers yield
an excitation parameter $U=r(n_e n_H)^{1/3} > 50$~pc~cm$^{-2}$, which serves as
a measure of the flux of ionizing photons from the central star.  This
corresponds to the excitation parameter expected for a zero age main sequence
O7 star with Lyman continuum flux $5\times10^{48}$ photons s$^{-1}$ and total
luminosity $> 10^5$~\Lsol\ \citep{Panagia1973}.  Since this is greater than the
luminosity of the entire Kleinmann-Low Nebula, we conclude that SrcI cannot be
an HII region.

Note that our analysis does not rule out some small contribution to the continuum
from free-free emission, as long as the recombination line intensities are
$<\,$5\% of the {\it total} continuum flux. 

\subsection{Electron-Neutral Free-Free Emission?}
\label{sec:Hminus}

\citet{Reid2007} considered the possibility that SrcI's continuum might
originate from electrons that are scattered by {\it neutral} hydrogen atoms or
molecules (the so-called H$^-$ and H$_2^-$ opacities).  This mechanism is
important for gas at temperatures between 1000 and 4500~K, in which Na, K, and
other metals are collisionally ionized but H and H$_2$ are mostly neutral.  It
is thought to explain the radio photospheres of Mira variables
\citep{Reid1997}.  Expressions for the H$^-$ and H$_2^-$ absorption
coefficients are given by \citet{Dalgarno1966}; \citet{Reid1997} provide
convenient analytical approximations to these expressions.  The
electron-neutral opacity is roughly 1000 times smaller than the electron-ion
opacity, so gas densities $>10^{11}$~cm$^{-3}$ typically are required to obtain
optically thick emission at mm wavelengths.  

As for regular p$^+$e$^-$ free-free emission, the H$^-$ and H$_2^-$ absorption
coefficients are proportional to $\nu^{-2}$, so at sufficiently high
frequencies the emission must become optically thin.  The spectral energy
distribution in Figure~\ref{fig:sed} shows no sign of flattening at higher
frequency, however.  Is it possible that in SrcI the electron-neutral emission
could be optically thick at frequencies as high as 660~GHz?  

\citet{Hirota2015} used the \citet{Reid1997} approximations to compute the
turnover frequency $\nu_{\rm to}$ where the (H$^-+$H$_2^-$) optical depth
becomes unity, for an 84 AU long path length comparable to the path length
through SrcI.  To obtain a turnover frequency above 600~GHz for gas at 1000~K
requires hydrogen density $n_{\rm H} = n($H + 2H$_2) > 3 \times
10^{15}$~cm$^{-3}$.  For a 20 AU thick $\times$ 84 AU diameter disk, this
corresponds to an absurdly high mass of $>$1000~\Msol.  The disk mass can be
much smaller if the gas is hotter, because the fractional ionization increases
steeply as the temperature rises.  For gas at 3000~K, the density is
$9\times10^{10}$~cm$^{-3}$ and the disk mass is only 0.035~\Msol.  Such a high
temperature is inconsistent, however, with both the continuum brightness
temperature and the SO$_2$ excitation temperature (Section~\ref{sec:rotdiag}),
both of which are $\lesssim 1000$~K.  For a disk with temperature below 1200~K
and a mass less than a few \Msol, Figure~A3 in \citet{Hirota2015} implies that
electron-neutral free-free emission cannot be a significant source of opacity
at frequencies higher than $\sim$200~GHz.  

\subsection{Dust?}
\label{sec:dust}

Thermal radiation from dust is the usual emission mechanism for circumstellar
disks at mm and submm wavelengths.  Generally the dust emission is assumed to be
optically thin, with a flux density that scales as 
\begin{equation} 
S_{dust}(\nu) \propto \kappa_{\nu}\,N\, B_{\nu}, 
\end{equation} 
where $\kappa_{\nu} \propto \nu^{\beta}$ is the absorption coefficient per unit
(gas+dust) mass, $N$ is the column density, and $B_{\nu} \propto \nu^2$ is the
Planck function.  Typically $\beta \sim 1$, so one expects the spectral energy
distribution to be steeper than $\nu^2$ at frequencies where dust emission is
dominant. 

Earlier measurements suggested that SrcI's spectral index did steepen at about
230~GHz, which led \citet{Beuther2006} and \citet{Hirota2015} to fit the SED
with a mixture of free-free emission, dominant below 230~GHz, and dust
emission, dominant above.  Our flux density measurements suggest
that the spectral index begins to steepen only above 350~GHz, however, and in
the sections above we argued that neither p$^+$e$^-$ nor electron-neutral
free-free emission could contribute significantly to the flux densities at
frequencies as high as 230~GHz.

Here we consider an alternative possibility, that SrcI's continuum emission at
mm and submm wavelengths originates from optically {\it thick} dust, consistent
with the observed $\nu^2$ spectrum.  We attribute the spectral steepening above
350~GHz to a thin, hot ``atmosphere'' on the outside of the disk that slightly
increases the source size or effective temperature at 660~GHz.  Some
contribution from H$^-$ or p$^+$e$^-$ free-free emission is required to explain
the continuum at low frequencies, where the dust emission must ultimately
become optically thin.   Free-free emission may also account for the central
point source that seems to be present in high resolution 43~GHz VLA maps of
SrcI \citep{Reid2007,Goddi2011}.  We will assume, however, that most of the
43~GHz continuum is attributable to dust.  \citet{Reid2007} argued that this
was unlikely on the grounds that molecular lines had not been detected toward
the disk, but ALMA data now show that lines are abundant at higher frequencies.

Assuming that the dust emission is optically thick at 43~GHz provides a
lower limit to the column density through the disk, and hence the disk mass.
We follow the procedure used by \citet{Chandler2005} to estimate the mass
opacity of dust $\kappa_{\rm 43\,GHz}$ (cross section per gram of gas plus
dust) at 43~GHz. They begin with the opacities $\kappa_{\rm ref}$ (cross
section per gram of refractory material) calculated by \citet{Ossenkopf1994}
for the case of coagulated grains with no ice mantles and a gas density $n_{\rm
H}$ = 10$^8$~cm$^{-3}$.  At 230~GHz, the lowest frequency tabulated,
$\kappa_{\rm ref} = 5.86$~cm$^2$ g$^{-1}$.  Extrapolating to 43~GHz assuming
$\beta$ = 1.1, and using a gas-to-dust mass ratio of 100, \citet{Chandler2005}
estimate $\kappa_{\rm 43\,GHz} = 9.3 \times 10^{-3}$~cm$^2$~g$^{-1}$.  To
obtain optically thick emission through the SrcI disk, we require that 
\begin{equation}
\tau_{\rm 43\,GHz} = \kappa_{\rm 43\,GHz}\, n_{\rm H}\, m_{\rm H}\, \mu\, D \gtrsim 1,
\label{eqn:tau} 
\end{equation} 
where $D \sim 100$~AU is the path length through the disk, $n_{\rm H} = n($H +
2H$_2)$ is the hydrogen density, $m_{\rm H}$ is the mass of a hydrogen atom,
and $\mu = 1.36$ is the ratio of the total gas mass to hydrogen mass.  Then we
find n$_{\rm H} \gtrsim 3.2\times 10^{10}$~cm$^{-3}$, and a disk mass
$\gtrsim$0.02~\Msol.     

The disk mass could be significantly larger if the dust grains in SrcI are very
small, or if they have grown to pebble-size.  For grains characteristic of the
diffuse interstellar medium, \citet{Ossenkopf1994} find $\kappa_{\rm ref} =
0.31$~cm$^2$ g$^{-1}$ at 230~GHz.  Extrapolating to 43~GHz using $\beta = 2$,
appropriate for very small grains, and again assuming a gas to dust mass ratio
of 100, one finds $\kappa_{\rm 43\,GHz} = 1.1 \times 10^{-4}$~cm$^2$.
Equation~\ref{eqn:tau} then implies a hydrogen density n$_{\rm H} =
2.7\times10^{12}$~cm$^{-3}$ and a disk mass of 1.7~\Msol.  A similar hydrogen
density and disk mass are obtained for pebble-sized (10 cm) grains, for which
$\kappa_{\nu} \sim 2\times10^{-4}$~cm$^2$~g$^{-1}$ for $\nu \gtrsim 43$~GHz
\citep{Testi2001}.  These mass estimates should be considered upper limits,
since they are based on unrealistically small grain opacities.  It's also
unlikely that gas densities exceed 10$^{11}$~cm$^{-3}$ on the disk surface, as
this would collisionally quench the SiO masers \citep{Lockett1992,Goddi2009}.
We conclude that the disk mass probably is in the range 0.02-0.2~\Msol, but
could be as large as 2~\Msol\ for more extreme dust properties.

As noted above, thermal emission from dust cannot explain the flux densities
measured below 43~GHz, many of which exceed the $\nu^2$ extrapolation from mm
wavelengths.  For example, the flux density of SrcI at 33.6~GHz is $\sim$10~mJy
\citep{Rivilla2015}, whereas optically thick dust would contribute only 7~mJy
at this frequency.   H$^-$ free-free emission is likely to be responsible
for the excess.  The dotted red curve in Figure~\ref{fig:sed} shows a free-free
spectrum with a turnover frequency of 30~GHz that is compatible with the low
frequency data.

The infrared spectrum of SrcI, glimpsed in light scattered off nearby
reflection nebulosities, indicates that the central object has a 3500-4500~K
photosphere \citep{Testi2010}.  Although H$^-$ free-free emission from this
photosphere must contribute to the continuum, it cannot fully explain the
low-frequency excess---even a 4000~K, 8~AU diameter source would contribute
less than 1~mJy to the 33.6~GHz flux density.  Free-free emission must
originate from a larger region, perhaps from the SiO maser zone along the disk
surface.  Our upper limit on the recombination line intensities also does not
rule out a small contribution from regular p$^+$e$^-$ free-free emission,
perhaps from shock-excited gas in the outflow.

\subsection{Source Model}
\label{sec:model}

The picture of SrcI that emerges is of a dusty disk or torus with a 
temperature of $\sim$500~K and a mass of $0.02-0.2$~\Msol\ that orbits a
5-7~\Msol\ central star (or binary).  A bipolar outflow is launched from the
surfaces of the disk.  These surfaces must be hotter (1200-1500~K) than the
interior of the disk in order to account for the SiO masers and other high
excitation lines that occur at the base of the outflow.  The high abundance of
\mbox{silicon-} and sulfur-bearing molecules in the atmosphere of the disk
suggests that dust grains are being destroyed in this layer.  In essence, SrcI
is an inside-out Mira variable: in a Mira, SiO masers occur just inside
the radius where dust condenses, whereas in SrcI the masers occur just 
outside the radius where dust grains are destroyed.

\subsection{Rethinking the Paradigm}

We now consider the implications of our data for models of star formation in
Orion.  Our results are not easily explained by scenarios in which SrcI
and BN were ejected by the dynamical decay of a multiple system 500 years ago.

Any model of the Orion-KL region must account for the fact that the
Becklin-Neugebauer Object, located just 10\arcsec\ NW of SrcI, is a runaway B
star \citep{Plambeck1995,Tan2004}.  VLA proper motion measurements
\citep{Goddi2011} show that it is moving to the NW at $26.4 \pm 2.1$~\kms\ in
the frame of the Orion Nebula Cluster.  BN's past trajectory passes close to
{\it both} SrcI and the Orion Trapezium.  

\citet{Tan2004} argued that BN was ejected from the Trapezium roughly 4000
years ago as a result of an interaction with $\theta^1$Ori~C, a compact binary
with a total mass of 45~\Msol.  \citet{Chatterjee2012} show that the observed
orbital parameters of $\theta^1$C closely match the values required to eject
BN, and argue that the likelihood that $\theta^1$C has these properties purely
by chance is $\sim 10^{-5}$.

Surprisingly, however, proper motion measurements indicate that SrcI is moving
to the SE at $11.5 \pm 2.1$~\kms\ \citep{Goddi2011}, almost directly away from
BN.  Tracing the BN and SrcI proper motion vectors backward, \citet{Goddi2011}
find that the projected separation of the 2 stars in the plane of the sky was
50~$\pm 100$~AU just 560 years ago.  This discovery led to the alternative
model in which SrcI and BN were ejected by the dynamical decay of a multiple
system \citep{Rodriguez2005,Gomez2008,Goddi2011}.

The dynamical decay model also provides an explanation for the system
of high velocity (up to 300~\kms) ``bullets,'' bow shocks and ``fingers'' that
are visible in lines of H$_2$, Fe II, and CO, and that appear to emerge from a
common center a few arcseconds NW of SrcI
\citep{Allen1993,Zapata2009,Bally2011,Bally2015}.  At one time the finger
system was assumed to be a high velocity outflow from SrcI, but now it is
recognized as the signpost of an explosive event 500-1000 years ago
\citep{Doi2002,Bally2011}.  The bullets are interpreted as dense fragments of
circumstellar disks that were torn apart in a stellar encounter and launched
into the surrounding cloud \citep{Bally2015}.  

Although the dynamical decay model has many attractive features,  two
observational results are difficult to understand:
\begin{enumerate} 
\item The mass of SrcI inferred from molecular line rotation
curves is only 5-7~\Msol, too small to have ejected BN with a speed of 25~\kms.
\item SrcI's 100~AU diameter, 0.1~\Msol\ disk is unlikely to have survived the
final stellar encounter, and it is too large and too massive to have accreted
in the 500 years since SrcI was ejected.
\end{enumerate} 
Let us consider these points more carefully.  First, is SrcI massive enough to
have ejected BN?  \citet{Goddi2011} did an extensive series of simulations to
test under what circumstances a 10~\Msol\ star can be ejected with speed $\geq
25$~\kms\ from a multiple system similar to the hypothesized BN-SrcI system.
The most favorable results were obtained if the initial system contained a
single 10~\Msol\ star and a preexisting (primordial) 20~\Msol\ binary.  In
about 40\% of these trials, dynamical interactions led to the ejection of both
objects, and a hardening of the binary orbit.  Sometimes the original binary
survived, and sometimes there was an exchange of stars.  The increase in
binding energy of the binary compensates for the kinetic energy of the runaway
stars.  Ejections were rare, however, if the initial system did not contain a
massive binary.  For example, only 1 of 1000 simulations that began with three
single 10~\Msol\ stars resulted in the ejection of a star with speed greater
than 25~\kms.  And, in simulations that began with an 10~\Msol\ star and an
11~\Msol\ binary (10+1 or 8+3~\Msol), only 1\% of the trials led to an
ejection, always of the lowest mass star, while the two more massive stars
formed a binary.  The conclusion is that the dynamical decay scenario 
strongly favors a mass $\gtrsim 20$~\Msol for SrcI.

Is it possible that the maser and molecular line rotation curves underestimate
the mass of SrcI by a factor of 3?  \citet{Matthews2010} and \citet{Goddi2011}
suggest that the measured mass should be taken as a lower limit, since SrcI's
circumstellar disk may be partially supported by magnetic fields or other
non-gravitational forces.  Much of the molecular line emission does seem to
originate from the base of the bipolar outflow, from gas that may not be
gravitationally bound to the central object.  However, the near infrared
spectrum of SrcI, which probes the gas velocity dispersion within a few AU of
the central object, also suggests a central mass of only $\sim 10$~\Msol\
\citep{Testi2010}, again lower than what is needed to eject BN. 

The dynamical ejection model also has difficulty explaining the existence of an
$\sim$0.1~\Msol\ disk around SrcI.  Either this is the remnant of a preexisting
disk that survived the close stellar encounter, or it has accumulated in the
last 500 years.  To explore the first possibility, \citet{Moeckel2012}
numerically simulated collisions between a 10~\Msol\ single star and a 13 or
20~\Msol\ (10+3 or 10+10~\Msol) binary with circumbinary disk.  Only a small
percentage of the trials led to outcomes in which more than 10\% of the disk
material was retained by the final binary, and in which the relative velocity
between the single star and the binary was $> 20$~\kms.

\citet{Bally2011} and \citet{Goddi2011} have discussed the other possibility,
that the disk accumulated in just 500 years via Bondi-Hoyle accretion
(gravitational focusing) as SrcI plowed through the ambient molecular cloud at
12~\kms.  If the mass of SrcI is 7~\Msol\, then the focusing radius $R_{BH}
\sim 40$~AU and the accumulated disk mass is a few~$\times~10^{-4}$~\Msol\ for
an ambient density n(H$_2$) $\sim$10$^7$~cm$^{-3}$.  This is two orders of
magnitude smaller than our estimate of the disk mass.  Furthermore, a disk
accumulated via gravitational focusing is expected to be irregular, with a
radius $\leq 0.1 \times R_{BH} = 4$~AU \citep{Goddi2011}, much smaller than the
observed disk.  The conclusion is that a disk as large and as massive as SrcI's
could not have formed by Bondi-Hoyle accretion, and is unlikely to be the
remnant of a preexisting disk that survived the dynamical encounter.

We conclude that the dynamical decay model is difficult to reconcile with the
observations, and that it is easier to explain the ejection of BN by an
interaction with $\theta^1$C than with SrcI. Unfortunately, this leaves
unexplained the large proper motion of SrcI, and the unique characteristics of
the finger system.  \citet{Chatterjee2012} suggested that the passage of
BN through the Kleinmann-Low star-forming core might have triggered enhanced
accretion and outflow activity from SrcI, but it is highly unlikely that BN
passed close enough to SrcI to accelerate it to a velocity of 10~\kms\ and to
eject the finger system.  Clearly, further investigations of this region are
needed in order to resolve these discrepancies.

\section{SUMMARY}
\label{sec:summary}

Orion SrcI was imaged with ALMA at 350 and 660 GHz with $\sim$0\farcs2 angular
resolution.  These results were compared with previous 229~GHz CARMA images
obtained with comparable angular resolution.  The goals were to infer the
nature of the continuum emission and to identify spectral lines from this
nearby high mass protostar.  The principal results are:

\begin{enumerate}

\item The continuum source has nearly the same size from 229 to 661~GHz,
approximately 0\farcs23 $\times$ 0\farcs07 (100 $\times$ 30~AU) according to
elliptical Gaussian fits.  It is interpreted as a circumstellar disk viewed
nearly edge-on.

\item The flux density of SrcI is proportional to $\nu^2$ from 43~GHz to
350~GHz, consistent with blackbody emission, then increases slightly faster
than $\nu^2$ from 350~GHz to 660~GHz.  The peak brightness temperature inferred
from the source size and flux density is $\sim$500~K. 

\item The H26$\alpha$ (353.6~GHz) and H21$\alpha$ (662.4~GHz) hydrogen
recombination lines were not detected.  The upper limit on the line intensities
of 0.05 times the continuum level rules out the possibility that SrcI is a
hypercompact HII region.   If such an HII region were optically thin, the
H21$\alpha$ line would be $\sim$5.7 times brighter than the free-free
continuum.  An optically thick HII region is ruled out because a star capable
of maintaining the ionization of such a dense region would have a luminosity
greater than that of the entire KL Nebula.

\item For gas temperatures $\lesssim 1000$~K implied by the source brightness
temperature, electron-neutral free-free emission (the H$^-$ opacity) can be
ruled out as a significant source of continuum emission at frequencies above
200~GHz, as it would require an implausibly high disk mass.

\item Optically thick thermal emission from dust is the most reasonable
explanation for the submm continuum.   The spectral energy distribution
suggests that the dust is optically thick down to frequencies of 43~GHz.  For a
gas-to-dust mass ratio of 100, and plausible dust opacity laws, we infer
a disk mass in the range 0.02 - 0.2~\Msol. 

\item Free-free emission, probably via the H$-$ mechanism, is needed to
explain the spectral energy distribution below 43~GHz, where the measured flux
densities exceed the $\nu^2$ extrapolation from mm wavelengths.  The free-free
emission is likely to originate from extended regions on the surfaces of the
disk, rather than from just a central point source.

\item A rich spectrum of molecular lines is observed toward SrcI.  Transitions
with upper energy levels $E_{\rm U} > 500$~K appear in emission and are
symmetric about \Vlsr = 5~\kms, while those with $E_{\rm U} < 500$~K appear as
blueshifted absorption features against the continuum, indicating that they
originate in outflowing gas.  High resolution images suggest that the emission
lines originate primarily from the ``atmosphere'' of the continuum disk, or the
base of the bipolar outflow.
 
\item Most of the emission lines are identified with \mbox{sulfur-} and
silicon--rich molecules (SO$_2$, SO, SiO, SiS).  In addition, there are several
unidentified emission lines; the most prominent of these is at 650.163~GHz.

\item Most of the emission lines have upper energy levels between 500 and
2000~K.  The most highly excited lines observed are v=1 J=3-2 CO (E$_{\rm U} =
3117$~K) and v=2 J=8-7 SiO (E$_{\rm U} = 3595$~K).  A rotational diagram
analysis of 16 SO$_2$ emission lines implies an excitation
temperature in the range 500-1000~K.

\item The emission lines exhibit a velocity gradient along the major axis of
the disk similar to that that measured previously for SiO masers
\citep{Matthews2010} and vibrationally excited H$_2$O \citep{Hirota2014},
consistent with rotation around a 5-7~\Msol\ star.

\end{enumerate}

We argue that SrcI is a dusty disk or torus that orbits a 5-7~\Msol\ central
object.  Dust grains are destroyed on the surface of the disk, in the zone
where SiO masers occur and where the bipolar outflow is launched.  The mass of
SrcI is a factor of 2-3 lower than predicted by the dynamical decay model
in which SrcI and BN are recoiling from one another after being ejected from a
multiple system.  The presence of a 100 AU diameter, 0.1~\Msol\ disk around the
source also is difficult to explain in this model.  

Additional measurements of SrcI's continuum spectral energy distribution
between 100 and 200~GHz would be useful to confirm that the flux densities
follow the expected $\nu^2$ law over this frequency range.  Higher angular
resolution observations also are needed to search for gaps or spiral arms
within the disk, to better constrain the continuum brightness temperature, and
to more accurately measure molecular line rotation curves.

\acknowledgments

This paper makes use of the following ALMA data:
ADS/JAO.ALMA\#2012.1.00123.S. ALMA is a partnership of ESO (representing
its member states), NSF (USA) and NINS (Japan), together with NRC
(Canada) and NSC and ASIAA (Taiwan), in cooperation with the Republic of
 Chile. The Joint ALMA Observatory is operated by ESO, AUI/NRAO and NAOJ.

The National Radio Astronomy Observatory is a facility of the National
Science Foundation operated under cooperative agreement by Associated
Universities, Inc."

Support for CARMA construction was derived from the states of California,
Illinois, and Maryland, the James S. McDonnell Foundation, the Gordon and Betty
Moore Foundation, the Kenneth T. and Eileen L. Norris Foundation, the
University of Chicago, the Associates of the California Institute of
Technology, and the National Science Foundation.

{\it Facilities:} \facility{CARMA}, \facility{ALMA}.
\bibliographystyle{apj}

\begin{thebibliography}{}
\expandafter\ifx\csname natexlab\endcsname\relax\def\natexlab#1{#1}\fi

\bibitem[{{Allen} \& {Burton}(1993)}]{Allen1993}
{Allen}, D.~A., \& {Burton}, M.~G. 1993, \nat, 363, 54

\bibitem[{{Bally} {et~al.}(2011){Bally}, {Cunningham}, {Moeckel}, {Burton},
  {Smith}, {Frank}, \& {Nordlund}}]{Bally2011}
{Bally}, J., {Cunningham}, N.~J., {Moeckel}, N., {et~al.} 2011, \apj, 727, 113

\bibitem[{{Bally} {et~al.}(2015){Bally}, {Ginsburg}, {Silvia}, \&
  {Youngblood}}]{Bally2015}
{Bally}, J., {Ginsburg}, A., {Silvia}, D., \& {Youngblood}, A. 2015, \aap, 579,
  A130

\bibitem[{{Beuther} {et~al.}(2006){Beuther}, {Zhang}, {Reid}, {Hunter},
  {Gurwell}, {Wilner}, {Zhao}, {Shinnaga}, {Keto}, {Ho}, {Moran}, \&
  {Liu}}]{Beuther2006}
{Beuther}, H., {Zhang}, Q., {Reid}, M.~J., {et~al.} 2006, \apj, 636, 323

\bibitem[{{Blake} {et~al.}(1987){Blake}, {Sutton}, {Masson}, \&
  {Phillips}}]{Blake1987}
{Blake}, G.~A., {Sutton}, E.~C., {Masson}, C.~R., \& {Phillips}, T.~G. 1987,
  \apj, 315, 621

\bibitem[{{Chandler} {et~al.}(2005){Chandler}, {Brogan}, {Shirley}, \&
  {Loinard}}]{Chandler2005}
{Chandler}, C.~J., {Brogan}, C.~L., {Shirley}, Y.~L., \& {Loinard}, L. 2005,
  \apj, 632, 371

\bibitem[{{Chatterjee} \& {Tan}(2012)}]{Chatterjee2012}
{Chatterjee}, S., \& {Tan}, J.~C. 2012, \apj, 754, 152

\bibitem[{{Dalgarno} \& {Lane}(1966)}]{Dalgarno1966}
{Dalgarno}, A., \& {Lane}, N.~F. 1966, \apj, 145, 623

\bibitem[{{Doi} {et~al.}(2002){Doi}, {O'Dell}, \& {Hartigan}}]{Doi2002}
{Doi}, T., {O'Dell}, C.~R., \& {Hartigan}, P. 2002, \aj, 124, 445

\bibitem[{{Forbrich} {et~al.}(2016){Forbrich}, {Rivilla}, {Menten}, {Reid},
  {Chandler}, {Rau}, {Bhatnagar}, {Wolk}, \& {Meingast}}]{Forbrich2016}
{Forbrich}, J., {Rivilla}, V.~M., {Menten}, K.~M., {et~al.} 2016, \apj, 822, 93

\bibitem[{{Genzel} {et~al.}(1981){Genzel}, {Reid}, {Moran}, \&
  {Downes}}]{Genzel1981}
{Genzel}, R., {Reid}, M.~J., {Moran}, J.~M., \& {Downes}, D. 1981, \apj, 244,
  884

\bibitem[{{Goddi} {et~al.}(2009){Goddi}, {Greenhill}, {Chandler}, {Humphreys},
  {Matthews}, \& {Gray}}]{Goddi2009}
{Goddi}, C., {Greenhill}, L.~J., {Chandler}, C.~J., {et~al.} 2009, \apj, 698,
  1165

\bibitem[{{Goddi} {et~al.}(2011){Goddi}, {Humphreys}, {Greenhill}, {Chandler},
  \& {Matthews}}]{Goddi2011}
{Goddi}, C., {Humphreys}, E.~M.~L., {Greenhill}, L.~J., {Chandler}, C.~J., \&
  {Matthews}, L.~D. 2011, \apj, 728, 15

\bibitem[{{G{\'o}mez} {et~al.}(2008){G{\'o}mez}, {Rodr{\'{\i}}guez}, {Loinard},
  {Lizano}, {Allen}, {Poveda}, \& {Menten}}]{Gomez2008}
{G{\'o}mez}, L., {Rodr{\'{\i}}guez}, L.~F., {Loinard}, L., {et~al.} 2008, \apj,
  685, 333

\bibitem[{{Greenhill} {et~al.}(2013){Greenhill}, {Goddi}, {Chandler},
  {Matthews}, \& {Humphreys}}]{Greenhill2013}
{Greenhill}, L.~J., {Goddi}, C., {Chandler}, C.~J., {Matthews}, L.~D., \&
  {Humphreys}, E.~M.~L. 2013, \apjl, 770, L32

\bibitem[{{Hirota} {et~al.}(2012){Hirota}, {Kim}, \& {Honma}}]{Hirota2012}
{Hirota}, T., {Kim}, M.~K., \& {Honma}, M. 2012, \apjl, 757, L1

\bibitem[{{Hirota} {et~al.}(2016){Hirota}, {Kim}, \& {Honma}}]{Hirota2016}
---. 2016, \apj, 817, 168

\bibitem[{{Hirota} {et~al.}(2014){Hirota}, {Kim}, {Kurono}, \&
  {Honma}}]{Hirota2014}
{Hirota}, T., {Kim}, M.~K., {Kurono}, Y., \& {Honma}, M. 2014, \apjl, 782, L28

\bibitem[{{Hirota} {et~al.}(2015){Hirota}, {Kim}, {Kurono}, \&
  {Honma}}]{Hirota2015}
---. 2015, \apj, 801, 82

\bibitem[{{Kim} {et~al.}(2008){Kim}, {Hirota}, {Honma}, {Kobayashi},
  {Bushimata}, {Choi}, {Imai}, {Iwadate}, {Jike}, {Kameno}, {Kameya},
  {Kamohara}, {Kan-Ya}, {Kawaguchi}, {Kuji}, {Kurayama}, {Manabe}, {Matsui},
  {Matsumoto}, {Miyaji}, {Nagayama}, {Nakagawa}, {Oh}, {Omodaka}, {Oyama},
  {Sakai}, {Sasao}, {Sato}, {Sato}, {Shibata}, {Tamura}, \&
  {Yamashita}}]{Kim2008}
{Kim}, M.~K., {Hirota}, T., {Honma}, M., {et~al.} 2008, \pasj, 60, 991

\bibitem[{{Lockett} \& {Elitzur}(1992)}]{Lockett1992}
{Lockett}, P., \& {Elitzur}, M. 1992, \apj, 399, 704

\bibitem[{{Matthews} {et~al.}(2010){Matthews}, {Greenhill}, {Goddi},
  {Chandler}, {Humphreys}, \& {Kunz}}]{Matthews2010}
{Matthews}, L.~D., {Greenhill}, L.~J., {Goddi}, C., {et~al.} 2010, \apj, 708,
  80

\bibitem[{{McKee} \& {Tan}(2003)}]{McKee2003}
{McKee}, C.~F., \& {Tan}, J.~C. 2003, \apj, 585, 850

\bibitem[{{Menten} {et~al.}(2007){Menten}, {Reid}, {Forbrich}, \&
  {Brunthaler}}]{Menten2007}
{Menten}, K.~M., {Reid}, M.~J., {Forbrich}, J., \& {Brunthaler}, A. 2007, \aap,
  474, 515

\bibitem[{{Moeckel} \& {Goddi}(2012)}]{Moeckel2012}
{Moeckel}, N., \& {Goddi}, C. 2012, \mnras, 419, 1390

\bibitem[{{Morino} {et~al.}(1998){Morino}, {Yamashita}, {Hasegawa}, \&
  {Nakano}}]{Morino1998}
{Morino}, J.-I., {Yamashita}, T., {Hasegawa}, T., \& {Nakano}, T. 1998, \nat,
  393, 340

\bibitem[{{M{\"u}ller} \& {Br{\"u}nken}(2005)}]{Muller2005}
{M{\"u}ller}, H.~S.~P., \& {Br{\"u}nken}, S. 2005, Journal of Molecular
  Spectroscopy, 232, 213

\bibitem[{{Niederhofer} {et~al.}(2012){Niederhofer}, {Humphreys}, \&
  {Goddi}}]{Niederhofer2012}
{Niederhofer}, F., {Humphreys}, E.~M.~L., \& {Goddi}, C. 2012, \aap, 548, A69

\bibitem[{{Ossenkopf} \& {Henning}(1994)}]{Ossenkopf1994}
{Ossenkopf}, V., \& {Henning}, T. 1994, \aap, 291, 943

\bibitem[{{Panagia}(1973)}]{Panagia1973}
{Panagia}, N. 1973, \aj, 78, 929

\bibitem[{{Plambeck} {et~al.}(1995){Plambeck}, {Wright}, {Mundy}, \&
  {Looney}}]{Plambeck1995}
{Plambeck}, R.~L., {Wright}, M.~C.~H., {Mundy}, L.~G., \& {Looney}, L.~W. 1995,
  \apjl, 455, L189+

\bibitem[{{Plambeck} {et~al.}(2009){Plambeck}, {Wright}, {Friedel}, {Widicus
  Weaver}, {Bolatto}, {Pound}, {Woody}, {Lamb}, \& {Scott}}]{Plambeck2009}
{Plambeck}, R.~L., {Wright}, M.~C.~H., {Friedel}, D.~N., {et~al.} 2009, \apjl,
  704, L25

\bibitem[{{Plambeck} {et~al.}(2013){Plambeck}, {Bolatto}, {Carpenter},
  {Eisner}, {Lamb}, {Leitch}, {Marrone}, {Muchovej}, {P{\'e}rez}, {Pound},
  {Teuben}, {Volgenau}, {Woody}, {Wright}, \& {Zauderer}}]{Plambeck2013}
{Plambeck}, R.~L., {Bolatto}, A.~D., {Carpenter}, J.~M., {et~al.} 2013, \apj,
  765, 40

\bibitem[{{Reid} \& {Menten}(1997)}]{Reid1997}
{Reid}, M.~J., \& {Menten}, K.~M. 1997, \apj, 476, 327

\bibitem[{{Reid} {et~al.}(2007){Reid}, {Menten}, {Greenhill}, \&
  {Chandler}}]{Reid2007}
{Reid}, M.~J., {Menten}, K.~M., {Greenhill}, L.~J., \& {Chandler}, C.~J. 2007,
  \apj, 664, 950

\bibitem[{{Rivilla} {et~al.}(2015){Rivilla}, {Chandler}, {Sanz-Forcada},
  {Jim{\'e}nez-Serra}, {Forbrich}, \& {Mart{\'{\i}}n-Pintado}}]{Rivilla2015}
{Rivilla}, V.~M., {Chandler}, C.~J., {Sanz-Forcada}, J., {et~al.} 2015, \apj,
  808, 146

\bibitem[{{Rodr{\'{\i}}guez} {et~al.}(2005){Rodr{\'{\i}}guez}, {Poveda},
  {Lizano}, \& {Allen}}]{Rodriguez2005}
{Rodr{\'{\i}}guez}, L.~F., {Poveda}, A., {Lizano}, S., \& {Allen}, C. 2005,
  \apjl, 627, L65

\bibitem[{{Rohlfs} \& {Wilson}(2000)}]{Rohlfs2000}
{Rohlfs}, K., \& {Wilson}, T.~L. 2000, {Tools of radio astronomy} (3rd edition;
  New York: Springer)

\bibitem[{{Schilke} {et~al.}(2001){Schilke}, {Benford}, {Hunter}, {Lis}, \&
  {Phillips}}]{Schilke2001}
{Schilke}, P., {Benford}, D.~J., {Hunter}, T.~R., {Lis}, D.~C., \& {Phillips},
  T.~G. 2001, \apjs, 132, 281

\bibitem[{{Schilke} {et~al.}(1997){Schilke}, {Groesbeck}, {Blake}, {Phillips},
  \& {T.~G.}}]{Schilke1997}
{Schilke}, P., {Groesbeck}, T.~D., {Blake}, G.~A., {Phillips}, \& {T.~G.} 1997,
  \apjs, 108, 301

\bibitem[{{Tan}(2004)}]{Tan2004}
{Tan}, J.~C. 2004, \apjl, 607, L47

\bibitem[{{Tang} {et~al.}(2010){Tang}, {Ho}, {Koch}, \& {Rao}}]{Tang2010}
{Tang}, Y.-W., {Ho}, P.~T.~P., {Koch}, P.~M., \& {Rao}, R. 2010, \apj, 717,
  1262

\bibitem[{{Testi} {et~al.}(2001){Testi}, {Natta}, {Shepherd}, \&
  {Wilner}}]{Testi2001}
{Testi}, L., {Natta}, A., {Shepherd}, D.~S., \& {Wilner}, D.~J. 2001, \apj,
  554, 1087

\bibitem[{{Testi} {et~al.}(2010){Testi}, {Tan}, \& {Palla}}]{Testi2010}
{Testi}, L., {Tan}, J.~C., \& {Palla}, F. 2010, \aap, 522, A44

\bibitem[{{Werner} {et~al.}(1976){Werner}, {Gatley}, {Becklin}, {Harper},
  {Loewenstein}, {Telesco}, \& {Thronson}}]{Werner1976}
{Werner}, M.~W., {Gatley}, I., {Becklin}, E.~E., {et~al.} 1976, \apj, 204, 420

\bibitem[{{Wright} {et~al.}(1995){Wright}, {Plambeck}, {Mundy}, \&
  {Looney}}]{Wright1995}
{Wright}, M.~C.~H., {Plambeck}, R.~L., {Mundy}, L.~G., \& {Looney}, L.~W. 1995,
  \apjl, 455, L185

\bibitem[{{Wu} {et~al.}(2014){Wu}, {Liu}, \& {Qin}}]{Wu2014}
{Wu}, Y., {Liu}, T., \& {Qin}, S.-L. 2014, \apj, 791, 123

\bibitem[{{Zapata} {et~al.}(2009){Zapata}, {Schmid-Burgk}, {Ho}, {Rodriguez},
  \& {Menten}}]{Zapata2009}
{Zapata}, L.~A., {Schmid-Burgk}, J., {Ho}, P.~T.~P., {Rodriguez}, L.~F., \&
  {Menten}, K. 2009, ArXiv e-prints, arXiv:0907.3945

\end{thebibliography}

\clearpage
\end{document}